\documentclass[fleqn,usenatbib]{mnras}
\usepackage{newtxtext,newtxmath}
\usepackage[T1]{fontenc}
\DeclareRobustCommand{\VAN}[3]{#2}
\let\VANthebibliography\thebibliography
\def\thebibliography{\DeclareRobustCommand{\VAN}[3]{##3}\VANthebibliography}

\usepackage{graphicx}	
\usepackage{amsmath}	

\title[Stellar/SMBH assembly in ULIRGs]{Stellar and black hole assembly in $z<0.3$ infrared-luminous mergers: intermittent starbursts vs. super-Eddington accretion}

\author[D. Farrah et al.]{
Duncan Farrah,$^{1,2}$\thanks{E-mail: dfarrah@hawaii.edu (DF)}
Andreas Efstathiou,$^{3}$
Jose Afonso,$^{4,5}$
Jeronimo Bernard-Salas,$^{6}$ 
Joe Cairns,$^{7}$ \newauthor
David L Clements,$^{7}$ 
Kevin Croker,$^{1}$  
Evanthia Hatziminaoglou,$^{8}$ 
Maya Joyce,$^{9}$ 
Mark Lacy,$^{10}$  \newauthor
Vianney Lebouteiller,$^{11}$  
Alix Lieblich,$^{1}$ 
Carol Lonsdale,$^{10}$ 
Seb Oliver,$^{12}$
Chris Pearson,$^{13,14,15}$  \newauthor
Sara Petty,$^{16,17}$   
Lura K Pitchford,$^{18,19}$ 
Dimitra Rigopoulou,$^{17}$ 
Michael Rowan-Robinson,$^{7}$
Jack Runburg,$^{1}$   \newauthor
Henrik Spoon,$^{20}$ 
Aprajita Verma,$^{17}$
and Lingyu Wang $^{21,22}$ 
\\
$^{1}$Department of Physics and Astronomy, University of Hawai'i, 2505 Correa Road, Honolulu, HI 96822, USA\\
$^{2}$Institute for Astronomy, 2680 Woodlawn Drive, University of Hawai'i, Honolulu, HI 96822, USA\\
$^{3}$School of Sciences, European University Cyprus, Diogenes Street, Engomi, 1516 Nicosia, Cyprus\\
$^{4}$Instituto de Astrof\'{i}sica e Ci\^{e}ncias do Espa\c co, Universidade de Lisboa, Portugal\\
$^{5}$Departamento de F\'{i}sica, Faculdade de Ci\^{e}ncias, Universidade de Lisboa, Portugal\\
$^{6}$ACRI-ST, Centre d’Etudes et de Recherche de Grasse (CERGA), 10 Av. Nicolas Copernic, 06130 Grasse, France\\
$^{7}$Imperial College London, Blackett Laboratory, Prince Consort Road, London, SW7 2AZ, UK\\
$^{8}$ESO, Karl-Schwarzschild-Str 2, D-85748 Garching bei München, Germany\\
$^{9}$Department of Physics and Astronomy, Michigan State University, East Lansing, MI 48824, USA\\
$^{10}$National Radio Astronomy Observatory, Charlottesville, VA, USA\\
$^{11}$AIM, CEA, CNRS, Universite Paris-Saclay, Universite Paris Diderot, Sorbonne Paris Cite, F-91191 Gif-sur-Yvette, France\\
$^{12}$Astronomy Centre, Department of Physics and Astronomy, University of Sussex, Brighton BN1 9QH, UK\\
$^{13}$RAL Space, STFC Rutherford Appleton Laboratory, Didcot, Oxfordshire OX11 0QX, UK\\
$^{14}$The Open University, Milton Keynes MK7 6AA, UK \\
$^{15}$Oxford Astrophysics, University of Oxford, Keble Rd, Oxford OX1 3RH, UK\\
$^{16}$NorthWest Research Associates, 3380 Mitchell Ln., Boulder, CO 80301, USA\\
$^{17}$Convent \& Stuart Hall Schools of the Sacred Heart, 2222 Broadway, San Francisco, CA 94115, USA\\
$^{18}$George P. and Cynthia Woods Mitchell Institute for Fundamental Physics and Astronomy, Texas A\&M University, College Station, TX, USA\\
$^{18}$Department of Physics and Astronomy, Texas A\&M University, College Station, TX, USA\\
$^{20}$Cornell Center for Astrophysics and Planetary Science, Space Sciences Building, Ithaca, NY 14853, USA\\
$^{21}$SRON Netherlands Institute for Space Research, Landleven 12, 9747 AD, Groningen, the Netherlands\\
$^{22}$Kapteyn Astronomical Institute, University of Groningen, Postbus 800, 9700 AV Groningen, the Netherlands
}

\date{Accepted 04/2022. Received 12/2021}

\pubyear{2022}

\begin{document}
\label{firstpage}
\pagerange{\pageref{firstpage}--\pageref{lastpage}}
\maketitle

\begin{abstract}
We study stellar and black hole mass assembly in a sample of 42 infrared-luminous galaxy mergers at $z<0.3$ by combining results from radiative transfer modelling with archival measures of molecular gas and black hole mass. The ratios of stellar mass, molecular gas mass, and black hole mass to each other are consistent with those of massive gas-rich galaxies at $z<0.3$.  The advanced mergers may show increased black hole mass to stellar mass ratios, consistent with the transition from AGN to ellipticals and implying substantial black hole mass growth over the course of the merger.  Star formation rates are enhanced relative to the local main sequence, by factors of $\sim100$ in the starburst and $\sim1.8$ in the host, respectively. The starburst star formation rates appear distinct to star formation in the main sequence at all redshifts up to at least $z\sim5$. Starbursts may prefer late-stage mergers, but are observed at any merger stage. We do not find evidence that the starbursts in these low-redshift systems substantially increase the total stellar mass, with a soft upper limit on the stellar mass increase from starburst activity of about a factor of two. In contrast, 12 objects show evidence for super-Eddington accretion, associated with late-stage mergers, suggesting that many AGN in infrared-luminous mergers go through a super-Eddington phase. The super-Eddington phase may increase black hole mass by up to an order of magnitude at an accretion efficiency of $42\pm33\%$ over a period of $44\pm22$\,Myr. Our results imply that super-Eddington accretion is an important black hole growth channel in infrared-luminous galaxies at all redshifts. 
\end{abstract}


\section{Introduction}
Mergers between gas-rich, massive galaxies play a role in galaxy assembly at nearly all redshifts. In the local universe, such mergers are rare, but almost invariably harbour one or both of rapid accretion onto a supermassive black hole (SMBH) and high rates of star formation  \citep{gen98,farrah03,arm06,nard09,psan21}. The inferred star formation and SMBH accretion rates make these systems viable sites for assembling significant stellar and black hole mass. Moreover, the merger itself may transform disk-like progenitors into elliptical remnants \citep{bar96,gen01}, with some fraction also passing through an optical quasar phase \citep{san88,tacc02,far09}. At higher redshifts the mechanisms behind high rates of star foramtion and SMBH accretion may be more diverse, but mergers remain important up to at least $z\sim6$, and may trigger even higher star formation and black hole accretion rates than their local counterparts \citep{alex05,pitch16,marr18,mrr18,gull19,wangl21,gao21}. Reviews of their properties are in \citet{sm96,blain02,lon06,cas14} and \citet{perez21}.

Tracking how star formation and AGN activity assemble stellar and SMBH mass in gas-rich mergers remains challenging. The star formation and AGN emission are almost always significantly obscured, rendering many X-ray through optical diagnostics unusable, and leading to high infrared luminosities, sometimes exceeding $10^{12}$L$_{\odot}$. It also remains uncertain if star formation and AGN activity can directly affect each other, and if so, in what sense (e.g. \citealt{fabian12,far12,harris16,pitch19,herr20,perna20}). Thus, even for low-redshift mergers it remains controversial when star formation and AGN are triggered over the course of the merger, how they are turned off, and how much stellar and black hole mass they assemble. 

In this paper we examine the relationships between star formation, AGN activity, stellar and SMBH mass assembly, and molecular gas in a sample of 42 gas-rich mergers with infrared luminosities in excess of $10^{12}$\,L$_{\odot}$ at $z<0.3$. We aim to gain insight into how molecular gas is converted to stellar and SMBH mass across the merger sequence. We do so by combining archival molecular gas, merger stage, and black hole mass estimates with new radiative transfer modelling which constrain the luminosities of star formation and AGN activity, as well as the stellar masses of the host galaxies.

Section \ref{sec:methods} describes the sample, the radiative transfer modelling results, the ancillary data, and the adopted comparison samples. In \S\ref{masscorrs} we examine the relations between stellar mass and both molecular gas mass and central black hole mass. In \S\ref{sec:starform} we quantify the relationship between stellar mass, and both starburst and host star formation rate (SFR), in context with the low-redshift star formation main sequence. We also study the relation between AGN luminosity and black hole mass. In \S\ref{massassemb} we study stellar and black hole mass assembly as a function of nuclear separation, to examine how mass in our sample is assembled across the merger sequence.  Further discussion is presented in \S\ref{hidpop}. Section \S\ref{sec:conclusions} presents our conclusions. In the appendix, we present analysis on some alternative explanations for the results on black hole mass assembly. We assume $L_{\odot} = 3.846\times10^{26}$\,W, \mbox{$H_0 = 70$\,km\,s$^{-1}$\,Mpc$^{-1}$}, \mbox{$\Omega = 1$}, and \mbox{$\Omega_{\Lambda} = 0.7$}. We convert all literature data to this cosmology where necessary.

\begin{table*}
\begin{center}
\begin{tabular}{cccccccccc}
\hline 
ID & IRAS       & Other       &Redshift& $\rm{M_{*}}$           & $\rm{M_{H_{2}}}$        & $\rm{M_{BH}}$ &  MS  & $n_{s}$ & Ref       \\ 
   & Name       & Names       &        & $10^{10}\rm{M_{\odot}}$ & $10^{9}\rm{M_{\odot}}$ & $10^{7}\rm{M_{\odot}}$ &      & Kpc       & $n_{s}$ \\ 
\hline 
1  & 00188-0856 & ---         & 0.128 & $22.3^{+2.9}_{-4.9}$  & $1.6\pm 0.2$ & ---   &$5$&Single & 2 \\ 
2  & 00397-1312 & ---         & 0.262 & $11.9^{+2.6}_{-3.3}$  & $34.4\pm11.2$& $1.1$ &$5$&Single & 2 \\ 
3  & 01003-2238 & ---         & 0.118 & $2.5^{+0.5}_{-0.3}$   & ---          & $2.5$ &$5$&Single & 3 \\ 
4  & 03158+4227 & ---         & 0.134 & $25.6^{+1.9}_{-3.6}$  & $6.6\pm 0.7$ & ---   &$2$&$42.80$& 10 \\ 
5  & 03521+0028 & ---         & 0.152 & $7.1^{+1.9}_{-0.6}$   & $7.8\pm 2.7$ & ---   &$3$&$3.88$ & 6 \\ 
6  & 05189-2524 & ---         & 0.043 & $14.5^{+2.7}_{-2.5}$  & $2.1\pm 0.2$ & $3.0$ &$5$&$0.19$ & 1 \\ 
7  & 06035-7102 & ---         & 0.079 & $9.5^{+0.9}_{-2.6}$   & $7.5\pm 1.5$ & $2.0$ &$2$&$10.40$& 7 \\ 
8  & 06206-6315 & ---         & 0.092 & $15.7^{+5.3}_{-4.2}$  & $16.4\pm 3.3$& ---   &$2$&$4.80$ & 4 \\ 
9  & 07598+6508 & ---         & 0.148 & $2.5^{+0.8}_{-0.3}$   & $10.8\pm 0.1$& ---   &$4$&Single & 2 \\ 
10 & 08311-2459 & ---         & 0.100 & $12.5^{+1.3}_{-1.1}$  & ---          & ---   &$4$&Single & 8 \\ 
11 & 08572+3915 & ---         & 0.058 & $2.4^{+0.4}_{-0.3}$   & $1.3\pm 0.2$ & ---   &$3$&$6.62$ & 1 \\ 
12 & 09022-3615 & ---         & 0.060 & $13.6^{+2.0}_{-1.6}$  & ---          & ---   &$3$&$4.03$ & 9 \\ 
13 & 10378+1109 & ---         & 0.136 & $8.6^{+4.1}_{-1.1}$   & ---          & ---   &$5$&Single & 6 \\ 
14 & 10565+2448 & ---         & 0.043 & $6.3^{+1.1}_{-0.3}$   & $5.3\pm 0.5$ & $2.0$ &$2$&$24.75$& 1 \\ 
15 & 11095-0238 & ---         & 0.107 & $2.1^{+0.3}_{-0.2}$   & $6.2\pm 2.0$ & $3.9$ &$4$&$1.29$ & 2 \\ 
16 & 12071-0444 & ---         & 0.128 & $10.0^{+1.8}_{-2.3}$  & ---          & $3.5$ &$3$&$2.80$ & 2 \\ 
17 & 13120-5453 & WKK2031     & 0.031 & $8.4^{+0.8}_{-1.2}$   & $5.4\pm 0.4$ & ---   &$5$&Single & 5 \\ 
18 & 13451+1232 & 4C12.50     & 0.122 & $25.2^{+ 18.2}_{-5.9}$& $7.6\pm 0.7$ & $6.5$ &$3$&$3.20$ & 3 \\ 
19 & 14348-1447 & ---         & 0.083 & $13.6^{+0.5}_{-2.8}$  & $13.6\pm 1.4$& $7.0$ &$3$&$5.47$ & 1 \\
20 & 14378-3651 & ---         & 0.068 & $4.5^{+0.6}_{-0.6}$   & $4.2\pm 0.8$ & $4.6$ &$5$&Single & 5 \\ 
21 & 15250+3609 & ---         & 0.055 & $2.8^{+0.2}_{-0.3}$   & $1.5\pm 0.5$ & $4.2$ &$4$&$1.27$ & 1 \\ 
22 & 15462-0450 & ---         & 0.100 & $3.6^{+3.0}_{-0.6}$   & $2.6\pm 0.2$ & $6.9$ &$4$&Single & 2 \\ 
23 & 16090-0139 & ---         & 0.134 & $9.4^{+1.1}_{-1.0}$   & $10.9\pm 3.6$& ---   &$4$&Single & 6 \\ 
24 & 17208-0014 & ---         & 0.043 & $5.9^{+1.6}_{-0.6}$   & $10.8\pm 1.1$&$23.3$ &$4$&Single & 5 \\ 
25 & 19254-7245 & SuperAntena & 0.062 & $63.2^{+25.0}_{-13.3}$& $8.2\pm 1.7$ & $7.9$ &$3$&$10.20$& 11 \\ 
26 & 19297-0406 & ---         & 0.086 & $11.7^{+2.9}_{-1.6}$  & $8.3\pm 0.3$ & ---   &$4$&$1.30$ & 4 \\ 
27 & 20087-0308 & ---         & 0.106 & $18.4^{+3.9}_{-5.7}$  & $14.3\pm 0.5$&$19.4$ &$4$&Single & 4 \\ 
28 & 20100-4156 & ---         & 0.130 & $8.7^{+5.1}_{-1.5}$   & $7.9\pm 0.6$ & ---   &$2$&$7.40$ & 4 \\ 
29 & 20414-1651 & ---         & 0.087 & $4.6^{+1.3}_{-0.6}$   & $2.8\pm 0.9$ &$10.3$ &$5$&Single & 4 \\ 
30 & 20551-4250 & ESO 286-19  & 0.043 & $5.5^{+1.1}_{-0.8}$   & $3.9\pm 0.8$ & $3.2$ &$4$&Single & 5 \\ 
31 & 22491-1808 & ---         & 0.078 & $5.7^{+0.7}_{-0.9}$   & $4.4\pm 0.5$ & $3.8$ &$3$&$2.68$ & 1 \\ 
32 & 23128-5919 & ESO 148-2   & 0.045 & $4.6^{+0.5}_{-0.6}$   & $3.4\pm 0.7$ & $4.4$ &$3$&$3.94$ & 4 \\ 
33 & 23230-6926 & ---         & 0.107 & $4.9^{+0.9}_{-1.0}$   & ---          & $3.5$ &$4$&$1.17$ & 4 \\ 
34 & 23253-5415 & AM 2325-541 & 0.130 & $17.8^{+3.8}_{-3.5}$  & ---          & ---   &$5$&Single & 7 \\ 
35 & 23365+3604 & ---         & 0.064 & $13.7^{+2.0}_{-1.1}$  & $6.0\pm 0.9$ & $3.7$ &$5$&Single & 5 \\ 
36 & 09320+6134 & UGC 5101    & 0.039 & $19.0^{+2.8}_{-1.7}$  & $4.9\pm 0.5$ &$55.0$ &$4$&$0.40$ & 1 \\ 
37 & 12540+5708 & Mrk 231     & 0.042 & $16.3^{+ 11.9}_{-3.8}$& $5.7\pm 1.0$ & $1.7$ &$4$&$0.64$ & 1 \\ 
38 & 13428+5608 & Mrk 273     & 0.037 & $12.6^{+1.5}_{-1.2}$  & $4.0\pm 0.4$ &$56.1$ &$4$&$0.77$ & 1 \\ 
39 & 13536+1836 & Mrk 463     & 0.049 & $3.8^{+4.2}_{-0.9}$   & $1.9\pm 0.6$ & $5.5$ &$3$&$4.40$ & 3 \\ 
40 & 15327+2340 & Arp 220     & 0.018 & $13.1^{+2.8}_{-1.0}$  & $5.1\pm 0.4$ & $6.1$ &$4$&$0.72$ & 1 \\ 
41 & 16504+0228 & NGC 6240    & 0.024 & $8.9^{+2.0}_{-1.8}$   & $7.0\pm 0.6$ &$120.0$&$4$&$0.89$ & 5 \\ 
42 & 01572+0009 & Mrk 1014    & 0.163 & $18.7^{+3.9}_{-5.3}$  & $5.7\pm 0.5$ & $13.5$&$4$&Single & 3 \\ 
\hline	
\end{tabular}
\caption{The sample and their basic properties. The stellar masses are taken from the radiative transfer model fits described in \S\ref{radfits} \citep{efs21}. The origins of the molecular gas and black hole masses are described in \S\ref{ancildata}. We assume a uniform uncertainty on the black hole masses of 30\%. The merger stages (MS) and projected nuclear separations ($n_{s}$) are described in \S\ref{ancildata}. The final column gives the origin of the nuclear separations: (1) \citealt{larson16}, (2) \citealt{vei06}, (3) \citealt{surace98}, (4) \citealt{bush02}, (5) \citealt{haan11} \& \citealt{kim13}, (6) \citealt{kim02}, (7) \citealt{rig99}, (8) \citealt{mur96}, (9) \citealt{rodr11}(10) \citealt{meus01}, (11) \citealt{colina91}
\label{tablethesample}}
\end{center}
\end{table*}

\begin{table*}
\begin{center}
\begin{tabular}{clllllllll}
\hline 
ID & \multicolumn{2}{c}{Star Formation Rate} &\multicolumn{2}{c}{CYGNUS AGN Luminosities}    & $\lambda_{e}$ & $\theta_{o}$ & \multicolumn{3}{c}{Other Corrected AGN Luminosities} \\	
   & Starburst  &  Host                      & Observed    & Corrected                &               &              & F06  & Si15  & St16  \\ 
   & \multicolumn{2}{c}{M$_{\odot}$yr$^{-1}$}&\multicolumn{2}{c}{$10^{12}$L$_{\odot}$}&               & $\degr$      & \multicolumn{3}{c}{$10^{12}$L$_{\odot}$} \\
\hline 
1 & $598^{+68}_{-121}$ & $2.0^{+1.5}_{-0.7}$ & $ 0.28^{+0.03}_{-0.05}$ & $ 1.21^{+0.22}_{-0.27}$ & --- & $ 47.0^{+1.4}_{- 1.3}$&  $ 1.23^{+0.36}_{-0.21}$ & $ 0.21^{+0.07}_{-0.06}$ & $ 0.17^{+0.20}_{-0.08 }$ \\ 
2 & $2192^{+691}_{-904}$ & $ 11.5^{+3.1}_{- 3.2}$ & $ 2.10^{+0.55}_{-0.31}$ & $ 47.79^{+22.09}_{-10.94}$ & $142.22^{+74.77}_{-48.26}$ & $ 73.5^{+0.0}_{- 0.4}$&  $ 43.96^{+6.28}_{-7.96}$ & $ 0.81^{+0.08}_{-0.19}$ & $ 80.78^{+34.20}_{-19.33 }$ \\ 
3 & $195^{+56}_{-78}$ & $2.9^{+0.7}_{- 0.4}$ & $ 0.81^{+0.22}_{-0.09}$ & $ 1.83^{+0.22}_{-0.25}$ & $ 2.25^{+0.63}_{-0.64}$ & $ 36.6^{+1.4}_{- 0.6}$&  $ 1.90^{+0.44}_{-0.47}$ & $ 1.12^{+0.15}_{-0.14}$ & $ 0.77^{+0.04}_{-0.10 }$ \\ 
4 & $660^{+174}_{-116}$ & $0.0^{+0.0}_{- 0.0}$ & $ 1.39^{+0.18}_{-0.23}$ & $ 7.62^{+1.08}_{-1.66}$ & --- & $ 37.9^{+1.2}_{- 1.9}$&  $ 1.74^{+0.15}_{-0.22}$ & $ 1.29^{+0.12}_{-0.29}$ & $ 0.07^{+0.01}_{-0.01 }$ \\ 
5 & $798^{+10}_{-39}$ & $0.0^{+0.1}_{- 0.0}$ & $ 0.26^{+0.04}_{-0.03}$ & $ 1.31^{+0.33}_{-0.34}$ & --- & $ 44.3^{+3.4}_{- 6.0}$&  $ 0.40^{+0.11}_{-0.11}$ & $ 0.51^{+0.06}_{-0.05}$ & $ 0.16^{+0.10}_{-0.06 }$ \\ 
6 & $170^{+56}_{-32}$ & $1.0^{+1.3}_{- 0.4}$ & $ 0.55^{+0.08}_{-0.06}$ & $ 4.71^{+0.60}_{-0.58}$ & $ 4.99^{+1.40}_{-1.39}$ & $ 60.0^{+0.4}_{- 1.6}$&  $ 0.86^{+0.14}_{-0.10}$ & $ 0.43^{+0.05}_{-0.14}$ & $ 0.20^{+0.05}_{-0.02 }$ \\ 
7 & $315^{+1237}_{-1218}$ & $ 11.7^{+0.8}_{- 4.9}$ & $ 0.21^{+0.09}_{-0.03}$ & $ 0.48^{+2.11}_{-0.10}$ & $ 0.74^{+3.24}_{-0.24}$ & $ 37.5^{+25.5}_{- 1.5}$&  $ 1.15^{+0.21}_{-0.23}$ & $ 0.13^{+0.04}_{-0.03}$ & $ 1.13^{+4.63}_{-1.10 }$ \\ 
8 & $378^{+95}_{-59}$ & $3.9^{+2.7}_{- 1.5}$ & $ 0.17^{+0.02}_{-0.02}$ & $ 0.54^{+0.13}_{-0.09}$ & --- & $ 41.4^{+1.4}_{- 3.8}$&  $ 0.51^{+0.10}_{-0.12}$ & $ 0.21^{+0.03}_{-0.04}$ & $ 0.16^{+0.03}_{-0.02 }$ \\ 
9 & $520^{+40}_{-27}$ & $0.0^{+0.1}_{- 0.0}$ & $ 3.72^{+0.33}_{-0.37}$ & $ 1.96^{+0.21}_{-0.20}$ & --- & $ 56.4^{+2.0}_{- 1.5}$&  $ 1.76^{+0.25}_{-0.35}$ & $ 4.14^{+1.01}_{-0.52}$ & $ 9.08^{+2.34}_{-2.25 }$ \\ 
10 & $683^{+50}_{-112}$ & $ 11.3^{+1.9}_{- 1.0}$ & $ 0.56^{+0.07}_{-0.08}$ & $ 2.51^{+0.32}_{-0.30}$ & --- & $ 51.0^{+1.8}_{- 0.7}$&  $ 2.20^{+0.32}_{-0.34}$ & $ 1.00^{+0.14}_{-0.40}$ & $ 0.65^{+0.11}_{-0.11 }$ \\ 
11 & $206^{+50}_{-66}$ & $1.6^{+0.7}_{- 0.2}$ & $ 0.78^{+0.15}_{-0.11}$ & $ 23.39^{+9.73}_{-4.27}$ & --- & $ 69.3^{+2.4}_{- 1.1}$&  $ 11.58^{+0.85}_{-1.92}$ & $ 0.25^{+0.07}_{-0.07}$ & $ 1.24^{+1.02}_{-0.40 }$ \\ 
12 & $482^{+20}_{-14}$ & $0.4^{+0.0}_{- 0.0}$ & $ 0.26^{+0.03}_{-0.02}$ & $ 3.80^{+0.68}_{-0.48}$ & --- & $ 73.3^{+0.2}_{- 0.4}$&  $ 0.08^{+0.01}_{-0.01}$ & $ 0.04^{+0.01}_{-0.02}$ & $ 0.13^{+0.02}_{-0.01 }$ \\ 
13 & $245^{+144}_{-121}$ & $5.3^{+2.8}_{- 3.5}$ & $ 0.83^{+0.12}_{-0.14}$ & $ 20.04^{+13.56}_{-7.75}$ & --- & $ 60.1^{+2.5}_{- 6.1}$&  $ 1.01^{+0.20}_{-0.20}$ & $ 0.44^{+0.07}_{-0.11}$ & $ 0.21^{+0.03}_{-0.04 }$ \\ 
14 & $322^{+21}_{-20}$ & $5.3^{+0.9}_{- 1.0}$ & $ 0.09^{+0.02}_{-0.01}$ & $ 0.22^{+0.05}_{-0.01}$ & $ 0.34^{+0.11}_{-0.09}$ & $ 39.0^{+2.6}_{- 0.6}$&  $ 0.03^{+0.01}_{-0.00}$ & $ 0.72^{+0.50}_{-0.12}$ & $ 0.12^{+0.02}_{-0.02 }$ \\ 
15 & $151^{+20}_{-30}$ & $1.8^{+0.4}_{- 0.2}$ & $ 0.94^{+0.12}_{-0.10}$ & $ 29.55^{+14.51}_{-6.45}$ & $23.56^{+12.39}_{-6.79}$ & $ 65.4^{+3.0}_{- 1.6}$&  $ 1.06^{+0.15}_{-0.16}$ & $ 0.49^{+0.07}_{-0.48}$ & $ 0.21^{+0.02}_{-0.03 }$ \\ 
16 & $298^{+185}_{-193}$ & $ 10.3^{+3.0}_{- 1.8}$ & $ 0.84^{+0.18}_{-0.10}$ & $ 3.34^{+1.53}_{-0.43}$ & $ 2.98^{+1.56}_{-0.84}$ & $ 38.9^{+3.4}_{- 1.7}$&  $ 1.82^{+0.34}_{-0.35}$ & $ 0.88^{+0.11}_{-0.13}$ & $ 0.75^{+0.10}_{-0.09 }$ \\ 
17 & $549^{+10}_{-12}$ & $8.6^{+1.1}_{- 1.7}$ & $ 0.10^{+0.02}_{-0.01}$ & $ 0.28^{+0.18}_{-0.05}$ & --- & $ 38.5^{+6.6}_{- 2.4}$&  $ 0.35^{+0.18}_{-0.16}$ & $ 0.14^{+0.03}_{-0.02}$ & $ 0.09^{+0.02}_{-0.01 }$ \\ 
18 & $87^{+44}_{-46}$ & $ 12.1^{+3.6}_{- 6.8}$ & $ 1.33^{+0.25}_{-0.22}$ & $ 15.48^{+2.61}_{-2.60}$ & $ 7.40^{+2.24}_{-2.23}$ & $ 60.0^{+0.5}_{- 1.9}$&  $ 1.43^{+0.40}_{-0.37}$ & $ 1.07^{+0.38}_{-0.14}$ & $ 0.90^{+0.29}_{-0.14 }$ \\ 
19 & $378^{+39}_{-10}$ & $ 16.1^{+0.7}_{- 3.6}$ & $ 0.39^{+0.04}_{-0.14}$ & $ 3.69^{+0.35}_{-1.77}$ & $ 1.64^{+0.47}_{-0.90}$ & $ 45.8^{+0.1}_{- 0.9}$&  $ 0.38^{+0.09}_{-0.27}$ & $ 0.26^{+0.05}_{-0.05}$ & $ 0.12^{+0.01}_{-0.02 }$ \\ 
20 & $347^{+7}_{-37}$ & $2.5^{+0.6}_{- 0.5}$ & $ 0.14^{+0.01}_{-0.01}$ & $ 0.81^{+0.16}_{-0.13}$ & $ 0.55^{+0.17}_{-0.16}$ & $ 48.9^{+2.4}_{- 3.1}$&  $ 0.30^{+0.07}_{-0.07}$ & $ 0.45^{+0.12}_{-0.09}$ & $ 0.09^{+0.01}_{-0.01 }$ \\ 
21 & $160^{+25}_{-18}$ & $2.7^{+0.3}_{- 0.4}$ & $ 0.37^{+0.04}_{-0.05}$ & $ 4.62^{+0.90}_{-0.78}$ & $ 3.40^{+1.07}_{-1.02}$ & $ 56.5^{+1.3}_{- 0.6}$&  $ 0.77^{+0.11}_{-0.10}$ & $ 0.41^{+0.04}_{-0.05}$ & $ 0.09^{+0.02}_{-0.01 }$ \\ 
22 & $255^{+91}_{-51}$ & $3.5^{+3.9}_{- 0.9}$ & $ 0.25^{+0.05}_{-0.04}$ & $ 0.30^{+0.06}_{-0.04}$ & $ 0.14^{+0.04}_{-0.04}$ & $ 45.0^{+1.4}_{- 1.5}$&  $ 0.23^{+0.04}_{-0.04}$ & $ 0.43^{+0.06}_{-0.05}$ & $ 0.13^{+0.04}_{-0.02 }$ \\ 
23 & $855^{+53}_{-89}$ & $4.2^{+1.4}_{- 0.9}$ & $ 0.26^{+0.07}_{-0.05}$ & $ 0.87^{+0.30}_{-0.18}$ & --- & $ 38.8^{+0.3}_{- 1.0}$&  $ 0.78^{+0.65}_{-0.21}$ & $ 0.18^{+0.02}_{-0.04}$ & $ 0.14^{+0.05}_{-0.03 }$ \\ 
24 & $768^{+9}_{-129}$ & $3.9^{+1.2}_{- 0.6}$ & $ 0.06^{+0.02}_{-0.01}$ & $ 0.16^{+0.10}_{-0.03}$ & $ 0.02^{+0.01}_{-0.01}$ & $ 36.1^{+5.2}_{- 0.1}$&  $ 0.12^{+0.06}_{-0.04}$ & $ 0.06^{+0.01}_{-0.02}$ & $ 0.06^{+0.01}_{-0.02 }$ \\ 
25 & $218^{+334}_{-529}$ & $3.0^{+0.8}_{- 0.8}$ & $ 0.28^{+0.20}_{-0.04}$ & $ 1.52^{+0.77}_{-0.25}$ & $ 0.60^{+0.34}_{-0.18}$ & $ 48.4^{+1.6}_{- 0.9}$&  $ 1.32^{+0.30}_{-0.20}$ & $ 0.38^{+0.19}_{-0.05}$ & $ 0.28^{+0.06}_{-0.14 }$ \\ 
26 & $625^{+119}_{-107}$ & $ 13.0^{+4.2}_{- 3.3}$ & $ 0.34^{+0.17}_{-0.05}$ & $ 4.24^{+13.24}_{-1.32}$ & --- & $ 57.0^{+10.5}_{- 5.2}$&  $ 0.57^{+0.12}_{-0.09}$ & $ 0.23^{+0.13}_{-0.05}$ & $ 0.17^{+0.03}_{-0.03 }$ \\ 
27 & $925^{+34}_{-149}$ & $2.6^{+2.9}_{- 2.6}$ & $ 0.10^{+0.03}_{-0.03}$ & $ 0.37^{+0.24}_{-0.18}$ & $ 0.06^{+0.04}_{-0.03}$ & $ 48.1^{+5.9}_{- 9.2}$&  $ 0.01^{+0.10}_{-0.00}$ & $ 0.02^{+0.03}_{-0.01}$ & $ 0.02^{+0.02}_{-0.00 }$ \\ 
28 & $772^{+268}_{-301}$ & $5.1^{+1.3}_{- 1.0}$ & $ 0.52^{+0.08}_{-0.10}$ & $ 2.51^{+0.52}_{-0.96}$ & --- & $ 44.9^{+1.0}_{- 8.9}$&  $ 1.84^{+0.29}_{-0.35}$ & $ 0.63^{+0.11}_{-0.13}$ & $ 1.20^{+0.17}_{-0.11 }$ \\ 
29 & $460^{+15}_{-67}$ & $4.2^{+1.9}_{- 0.7}$ & $ 0.15^{+0.03}_{-0.04}$ & $ 1.49^{+0.51}_{-0.79}$ & $ 0.45^{+0.19}_{-0.27}$ & $ 54.0^{+5.3}_{- 7.6}$&  $ 0.06^{+0.03}_{-0.02}$ & $ 0.08^{+0.02}_{-0.01}$ & $ 0.08^{+0.02}_{-0.01 }$ \\ 
30 & $150^{+77}_{-56}$ & $6.6^{+1.5}_{- 1.5}$ & $ 0.31^{+0.06}_{-0.03}$ & $ 1.59^{+0.37}_{-0.18}$ & $ 1.54^{+0.53}_{-0.42}$ & $ 41.1^{+1.9}_{- 0.3}$&  $ 0.80^{+0.07}_{-0.10}$ & $ 0.27^{+0.05}_{-0.04}$ & $ 0.04^{+0.01}_{-0.01 }$ \\ 
31 & $235^{+12}_{-18}$ & $6.4^{+1.0}_{- 1.5}$ & $ 0.36^{+0.03}_{-0.06}$ & $ 5.82^{+1.42}_{-1.71}$ & $ 4.78^{+1.67}_{-1.84}$ & $ 55.8^{+3.8}_{- 4.3}$&  $ 0.38^{+0.07}_{-0.07}$ & $ 1.24^{+0.15}_{-0.05}$ & $ 0.11^{+0.02}_{-0.01 }$ \\ 
32 & $165^{+21}_{-26}$ & $4.9^{+0.7}_{- 1.4}$ & $ 0.13^{+0.01}_{-0.02}$ & $ 0.37^{+0.04}_{-0.06}$ & $ 0.26^{+0.07}_{-0.08}$ & $ 38.6^{+2.1}_{- 1.0}$&  $ 0.38^{+0.05}_{-0.08}$ & $ 0.18^{+0.04}_{-0.05}$ & $ 0.15^{+0.03}_{-0.02 }$ \\ 
33 & $364^{+15}_{-43}$ & $1.7^{+0.3}_{- 0.3}$ & $ 0.34^{+0.05}_{-0.04}$ & $ 1.57^{+0.43}_{-0.14}$ & $ 1.40^{+0.52}_{-0.37}$ & $ 36.8^{+2.1}_{- 0.4}$&  $ 0.57^{+0.14}_{-0.08}$ & $ 0.36^{+0.03}_{-0.08}$ & $ 0.12^{+0.02}_{-0.04 }$ \\ 
34 & $275^{+136}_{-128}$ & $ 21.0^{+4.3}_{- 6.1}$ & $ 0.57^{+0.12}_{-0.17}$ & $ 6.40^{+3.64}_{-3.42}$ & --- & $ 50.1^{+6.5}_{- 5.9}$&  $ 0.54^{+0.17}_{-0.24}$ & $ 1.50^{+0.25}_{-0.35}$ & $ 0.15^{+0.04}_{-0.05 }$ \\ 
35 & $375^{+41}_{-71}$ & $0.5^{+0.1}_{- 0.1}$ & $ 0.26^{+0.07}_{-0.03}$ & $ 1.57^{+1.21}_{-0.42}$ & $ 1.32^{+1.07}_{-0.49}$ & $ 41.7^{+5.7}_{- 3.8}$&  $ 0.30^{+0.08}_{-0.07}$ & $ 1.03^{+0.18}_{-0.22}$ & $ 0.08^{+0.01}_{-0.01 }$ \\ 
36 & $254^{+19}_{-26}$ & $1.1^{+0.2}_{- 0.2}$ & $ 0.09^{+0.02}_{-0.01}$ & $ 0.36^{+0.17}_{-0.11}$ & $ 0.02^{+0.01}_{-0.01}$ & $ 48.7^{+7.6}_{- 8.1}$&  $ 0.10^{+0.02}_{-0.01}$ & $ 0.01^{+0.00}_{-0.00}$ & $ 0.04^{+0.68}_{-0.01 }$ \\ 
37 & $518^{+114}_{-128}$ & $7.6^{+3.9}_{- 4.1}$ & $ 1.46^{+0.17}_{-0.13}$ & $ 5.98^{+0.70}_{-0.50}$ & $10.80^{+2.98}_{-2.85}$ & $ 55.2^{+1.5}_{- 1.4}$&  $ 3.09^{+0.35}_{-0.28}$ & $ 3.31^{+0.23}_{-0.55}$ & $ 0.79^{+0.14}_{-0.11 }$ \\ 
38 & $373^{+31}_{-71}$ & $0.5^{+0.3}_{- 0.1}$ & $ 0.17^{+0.04}_{-0.03}$ & $ 0.78^{+0.67}_{-0.32}$ & $ 0.04^{+0.04}_{-0.02}$ & $ 44.5^{+7.9}_{- 8.5}$&  $ 0.51^{+0.11}_{-0.07}$ & $ 0.42^{+0.31}_{-0.19}$ & $ 0.13^{+0.28}_{-0.02 }$ \\ 
39 & $54^{+19}_{-14}$ & $4.7^{+1.1}_{- 3.7}$ & $ 0.52^{+0.07}_{-0.05}$ & $ 0.52^{+0.10}_{-0.05}$ & $ 0.29^{+0.11}_{-0.10}$ & $ 47.2^{+0.3}_{- 1.4}$&  $ 0.71^{+0.10}_{-0.07}$ & $ 1.88^{+0.08}_{-0.09}$ & $ 1.60^{+0.19}_{-0.23 }$ \\ 
40 & $274^{+75}_{-55}$ & $0.5^{+0.2}_{- 0.1}$ & $ 0.12^{+0.08}_{-0.01}$ & $ 0.97^{+1.49}_{-0.15}$ & $ 0.50^{+0.77}_{-0.15}$ & $ 45.9^{+5.5}_{- 2.0}$&  $ 0.06^{+0.02}_{-0.01}$ & $ 0.53^{+0.02}_{-0.12}$ & $ 0.04^{+0.03}_{-0.00 }$ \\ 
41 & $107^{+69}_{-37}$ & $7.2^{+2.3}_{- 3.2}$ & $ 0.12^{+0.02}_{-0.03}$ & $ 0.47^{+0.06}_{-0.19}$ & $ 0.01^{+0.00}_{-0.01}$ & $ 40.5^{+1.1}_{- 4.4}$&  $ 0.38^{+0.09}_{-0.06}$ & $ 0.16^{+0.05}_{-0.03}$ & $ 0.02^{+0.00}_{-0.00 }$ \\ 
42 & $711^{+132}_{-141}$ & $ 23.5^{+5.4}_{- 7.2}$ & $ 1.29^{+0.20}_{-0.15}$ & $ 0.57^{+0.10}_{-0.08}$ & $ 0.13^{+0.04}_{-0.04}$ & $ 45.0^{+3.6}_{- 8.8}$&  $ 0.82^{+0.10}_{-0.08}$ & $ 1.36^{+0.15}_{-0.17}$ & $ 2.16^{+0.52}_{-0.26 }$ \\ 
\hline	
\end{tabular}
\caption{A summary of the SFRs, bolometric luminosities, and torus half-opening angles ($\theta_{o}$) used in this paper (\S\ref{radfits}). The CYGNUS luminosities are used in the main text, while the \citet{fritz06}, \citet{sieben15}, and \citet{stal16} luminosities are used in the appendix. All three alternative AGN models also feature anisotropic emission, but for brevity we present here only their corrected luminosities. The observed luminosities are given in EFS21. The Eddington ratios are calculated using the black hole masses from Table 1 and the anisotropy-corrected CYGNUS AGN luminosities. Further discussion on the Eddington ratios is given in \S\ref{bhmass}.\label{tablelums}}
\end{center}
\end{table*}

\section{Methods}\label{sec:methods}

\subsection{Sample Selection}
The sample is comprised of the 42 ultraluminous infrared galaxies (ULIRGs, systems with rest-frame $1-1000\mu$m luminosities in excess of $10^{12}$L$_{\odot}$) observed by the HERschel ULIRG Reference Survey (HERUS) carried out by the {\itshape Herschel} Space Observatory \citep{pilb10}. A complete description of the selection criteria of the sample is in \citet{far13,spoon13,pea16} and \citet{cle18}, so we summarize the selection here. The sample include all ULIRGs with IRAS 60$\mu$m fluxes greater than $\sim$2Jy, and includes all known ULIRGs at $z<0.3$. The sample is thus representative of infrared-luminous mergers in the low-redshift universe, and benefits from several decades of study from the X-ray to the radio. We present the sample in Table \ref{tablethesample}.

\subsection{Radiative Transfer Models}\label{radfits}
Star formation, AGN, and host galaxy properties are taken from the radiative transfer modelling study of \citet{efs21}. This study (hereafter EFS21) simultaneously fits separate component models for starbursts, AGN, and the host galaxy to all 42 objects in our sample. Their approach does not assume energy balance, and infers luminosities that are highly consistent with those inferred from other Spectral Energy Distribution (SED) models, or from e.g. emission line based calibrations. EFS21 also perform a thorough comparison between four different AGN obscurer models, both in terms of quality of fit and differences in physical parameters, and quantify the impact of anisotropic emission on the AGN luminosities. 

We adopt the CYGNUS model fits from EFS21 as in most cases they provide the best fit. These models have been tested successfully in previous studies \citep{lon15,mat18,herr17,kool20}, to separate the emission from accreting SMBHs, star formation, and the host galaxy. The data we use from EFS21 are:

\begin{itemize}
\item $\dot{M}_{Sb}$, $\dot{M}_{*}$: SFRs from the starburst and host component model fits. The starburst SFR is averaged over the age of the starburst and does not include the host galaxy SFR. 

\item $L^{o}_{AGN}$ \& $L^{c}_{AGN}$: Observed and anisotropy-corrected bolometric luminosity of the AGN. The anisotropy correction arises from the axisymmetric structure of the AGN obscurer and is discussed in EFS21. 

\item $\theta_{o}$: Half-opening angle of the AGN obscurer, measured from pole-on. 

\item $M_{*}$: Total stellar mass of the host galaxy.

\end{itemize}

\noindent The AGN model \citep{efsrr95,efstathiou95,efstathiou13} describes the obscurer with a dusty, smooth, tapered disk in a $r^{-1}$ density distribution. The model parameters include half-opening angle of the torus, inclination angle of the torus relative to the observer, ratio of inner to outer radius and equatorial optical depth. The starburst model \citep{efstathiou09} incorporates stellar population synthesis and allows for star-forming clouds to have different evolutionary stages. The model parameters are: the age of the starburst, initial optical depth, and time constant of the exponentially decaying SFR. The host models use a star formation history in which the stars and dust are mixed in a S\'{e}rsic profile. The model assumes three main parameters: the e-folding time of the star formation, the optical depth of the galaxy, and the ratio of the central intensity of starlight to that in the solar neighborhood. 

We present the stellar masses and SFRs from these fits in Tables \ref{tablethesample} and  \ref{tablelums} respectively, and the AGN luminosities in Table \ref{tablelums}.

Finally, in the appendix, for comparison with the AGN luminosities from CYGNUS, we use model fits with three alternative AGN models; those from  \citet{fritz06}, \citet{sieben15}, and \citet{stal16}. The models and resuts are described in detail in EFS21. We summarize the luminosities used here in Table \ref{tablelums}.

\subsection{Ancillary Data}\label{ancildata}

\subsubsection{Merger Stages}
To distinguish between early- and late-stage mergers, we employ merger classifications, primarily from \citet{larson16}, with additional data from \citet{colina91,mur96,surace98,rig99,meus01,bush02,kim02,vei06,rodr11} and \citet{haan11}. We convert other merger classifications to the Larson system for consistency. All of the sample are consistent with a classification within `Major Merger' stages 2-5.

When more fine-grained estimates of merger stage are useful, we employ projected nuclear separations, assembled from \citet{colina91,mur96,surace98,rig99,meus01,bush02,kim02,vei06,rodr11} and \citet{haan11}. Wherever possible we use high-resolution near-infrared imaging to measure nuclear separations as they are less affected by differential obscuration than optical imaging. The merger stages and nuclear separations are in nearly all cases consistent, with a boundary for $MS\geq4$ at $2$\,kpc.

\subsubsection{Molecular Gas Masses}
We adopt the luminosity of the J=1-0 transition of Carbon Monoxide (hereafter L$_{CO}$) as the tracer of the total molecular gas reservoir in our sample. Observations of L$_{CO}$ are available for most of our sample from \citet{kamen16}, including data from \citet{san89,mira90,downes93,solom97,baan08,chung09,grev09,braun11,papa12,xia12,meij13,ueda14,mash15,sliwa17,gowa18,ruff18,brown19,herr19,foto19}, and \citet{tan21}. We convert from $S_{CO}\Delta v$ in Jy km s$^{-1}$ to molecular gas mass via \citep{solv05}:

\begin{equation}
M_{H_{2}} =  \alpha_{CO} \frac{3.25\times 10^{7}S_{CO}\Delta v D_{L}^{2}}
{\nu_{obs}^{2} (1 + z)^{3} }
 \end{equation}

\noindent where $\nu_{obs}$ is in GHz and $D_{L}$ is in Mpc. We assume $\alpha_{CO} = 0.8$ (though see e.g. \citealt{carl17,herr19}).

\subsubsection{Black Hole Masses}
We assemble dynamically measured black hole masses ($M_{BH}$) from the studies of \citet{dasyra06,kawa07,wang07} and \citet{medling15}. We adopt dynamical, rather than virial or photometric masses, and discuss this choice in \S\ref{appbhmass}. For systems with mass measurements of two black holes,  we adopt only the mass for the infrared-brightest nucleus, and do not sum the black hole masses.

We assume that the anisotropy-corrected bolometric AGN luminosity is related to the accretion rate via:

\begin{equation}
L^{c}_{AGN} = \epsilon \dot{M}_{BH} c^{2} 
\end{equation}

\noindent where $\epsilon$ is the fraction of gravitational potential energy converted to luminosity during accretion (see also \S\ref{bhmass}). 

We compute Eddington luminosities ($L_{e}$) from the black hole masses assuming pure hydrogen ionization with opacity from Thomson scattering:

\begin{equation}
L_{e} = \frac{4 \pi G M_{BH} m_{p} c}{\sigma_{T}} \simeq 3.2\times10^{4}M_{BH}
\end{equation}
	
\noindent and then evaluate:

\begin{equation}
\lambda_{e} = \frac{L^{c}_{AGN}}{L_{e}}
\end{equation}

\noindent to obtain Eddington ratios ($\lambda_{e}$).

\subsubsection{Incompletenesses in Ancillary Data}\label{incans}
The sample is incomplete in both molecular gas masses and black hole masses. Seven objects do not have L$_{CO}$ data. Fourteen objects do not have dynamical M$_{BH}$ masses. The objects with missing data are however similar in terms of their other properties to the sample with complete data. We therefore do not believe that these omissions significantly bias our results. In the following we refer to the sample as a single entity, with a reduction in numbers in the appropriate plots. The potential for sample bias is discussed further in \S\ref{abias}.

\subsection{Comparison Samples}\label{compsamp}
There is no single comparison sample that can set our sample within the broader context of AGN and quiescent galaxies at low redshift. We therefore employ four distinct comparison samples, depending on the analysis being performed.

First is the xCOLDGASS sample (\citealt{sai17,cat18}, xCG hereafter). This survey includes CO J=1-0 observations with the IRAM 30m telescope of 532 galaxies at low redshift, selected from the Sloan Digital Sky Survey, and is designed to be a reference survey for the cold molecular gas properties of galaxies in the nearby universe. It is also used in studies of the redshift evolution of gas scaling relations (e.g. \citealt{tacc18}). We use the xCG sample as a comparison for stellar masses, molecular gas masses, and host SFRs in the low redshift galaxy population. 

Second is the Close AGN Reference Survey (\citealt{huse22,smir22}, CARS hereafter). The CARS sample consists of 40 moderately luminous, unobscured AGN in the redshift range $0.01<z<0.06$. CARS includes molecular gas masses, total star formation rates and stellar masses, black hole masses, and AGN luminosities. It is designed to be an anchor point for surveys of AGN at high redshift. We exploit the CARS sample as a comparison for stellar masses, molecular gas masses, total SFRs, black hole masses, and AGN luminosities in low-redshift AGN, but with the caveat that the CARS AGN luminosities are on average much lower than in our sample. 

The final two samples are both of luminous AGN, drawn from the Sloan Digital Sky Survey. First are the type 1 AGN from the SDSS data release 14 (\citealt{raks20}, see also \citealt{shen11,koz17}). Restricting to those quasars in the same redshift range as our sample and with a reliable single-epoch virial black hole mass gives 2106 objects. Second are the type 2 AGN presented by \citet{kong18}, again selected from the SDSS (see also \citealt{reyes08}). After restricting to objects in the same redshift range as our sample, this gives 369 objects. We use these two samples as comparisons for AGN luminosities, black hole masses, and Eddington ratios (but not SFRs or molecular gas masses) in AGN in comparable redshift and AGN luminosity ranges to our sample. 
 
We note that the three comparison AGN samples are optically, rather than infrared, selected. Their black hole masses are measured using the single-epoch virial approach, rather than stellar dynamics. These factors mean that the comparisons presented in the following should be interpreted with caution.

\begin{figure*}
\begin{center}
\includegraphics[width=14cm]{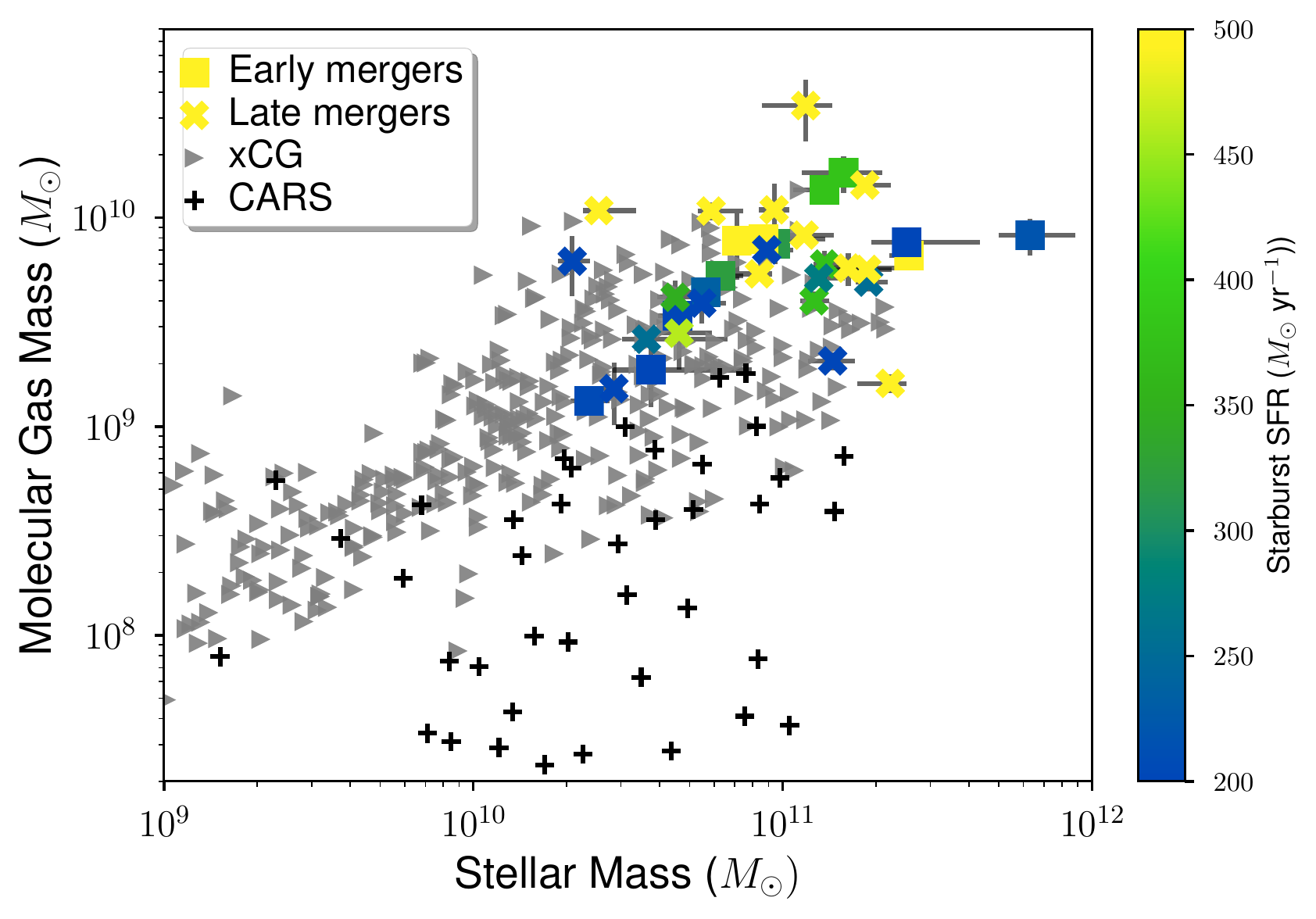}
\includegraphics[width=14cm]{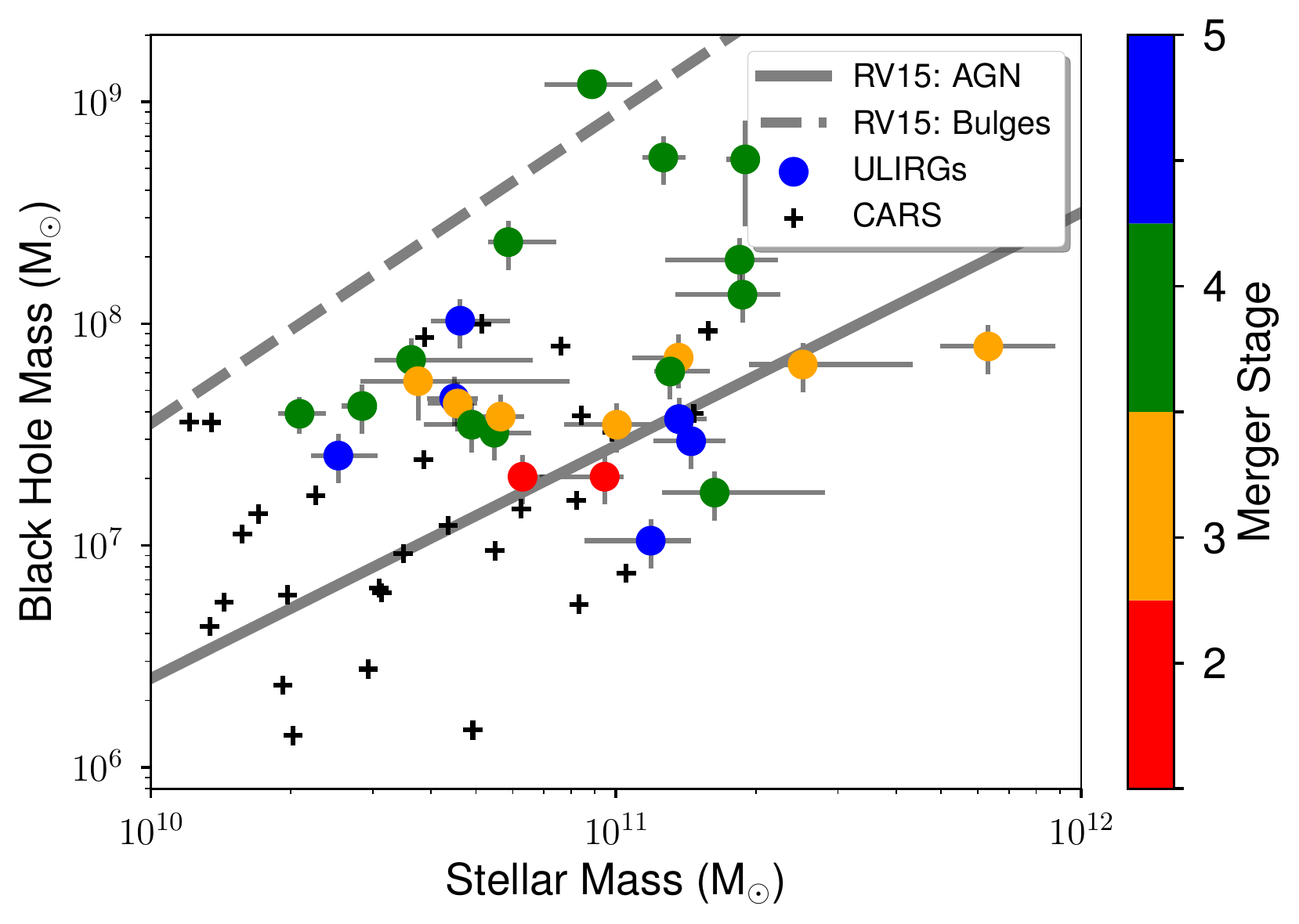}
\caption{{\itshape Top:} The $\rm{M_{*} - \rm{M_{H_{2}}}}$ distribution of our sample (\S\ref{masscorrsgas}). The starburst SFRs may be enhanced in more gas-rich systems. Comparing to the xCG and CARS samples (\S\ref{compsamp}) infers that our sample are typical of gas-rich, massive galaxies in the local universe and have higher molecular gas masses than lower luminosity AGN. {\itshape Bottom:} the $\rm{M_{*} - \rm{M_{BH}}}$ distribution of our sample (\S\ref{masscorrsbh}). The sample are mostly distributed between the local AGN and bulge relations in \citep{rein15}, and show tentative evidence for different distributions in the early vs. late-stage mergers. Compared to lower luminosity AGN in the same redshift range from CARS, our sample have higher black hole masses, and slightly higher total stellar masses.}
\label{fig:mstarmbh}
\end{center}
\end{figure*}

\section{Mass correlations}\label{masscorrs}
We here explore the assembly of stellar and black hole mass in our sample by considering two correlations with total stellar mass. In \S\ref{masscorrsgas} we examine the stellar mass vs. molecular gas mass relation. In \S\ref{masscorrsbh} we consider the stellar mass vs. black hole mass relation.

\subsection{Stellar Mass vs. Molecular Gas Mass}\label{masscorrsgas}
Stellar mass broadly scales with molecular gas mass in our sample (Figure \ref{fig:mstarmbh}, top). The Kendall Tau coefficient and p-value are ($\tau$, $p$) = (0.27, 0.02). Parametrizing the $\log M_{H_{2}} - \log M_{*}$ distribution as a linear relation\footnote{All parametrized fits were performed with the Orthogonal Distance Regression algorithm \citep{bog87} as implemented within the {\itshape SciPy} Python library and a random seed of 2001.}, we find:

\begin{equation}
\log \left(\frac{M_{H_{2}}}{M_{\odot}}\right) = (1.24 \pm 0.32)\left(\frac{M_{*}}{M_{\odot}}\right) - (3.89 \pm 3.47)
\end{equation}

\noindent Systems with higher starburst SFRs may prefer higher gas fractions. Compared to the xCG sample, most of our sample are gas-rich, but not exceptionally so - they seem to be `normal' gas-rich, massive galaxies at low redshift. Compared to the CARS AGN sample, our sample typically have higher stellar and molecular gas masses, consistent with more luminous activity in more gas-rich systems. 

We observe no relation in Figure \ref{fig:mstarmbh} top with merger stage; later-stage (MS$=4-5$) systems are not unusually gas-rich or gas-poor compared to early-stage (MS$=2-3$) mergers. This suggests that the molecular gas is not systematically converted to stellar mass over the entirety of the merger, though depletion during the latter merger stages is possible.

\subsection{Stellar Mass vs. Black Hole Mass}\label{masscorrsbh}
For the whole sample there is no evidence for a correlation between $M_{*}$ and $M_{BH}$ (Figure \ref{fig:mstarmbh} ($\tau$, $p$) = (0.12, 0.38)). The sample mostly lie between the $M_{*} - M_{BH}$ relations for low-redshift AGN and bulges (\citealt{rein15}, though see also \citealt{benn15,benn21}). Most are closer to the AGN relation, though there is no strong tendency for the more luminous AGN to prefer proximity to the AGN line (see also \citealt{ima20}). Compared to the CARS AGN sample, our sample typically have more massive central black holes. 

The early- and late-stage mergers show marginal evidence for different distributions in the $M_{*} - M_{BH}$ plane (two-sided KS test results of (0.41, 0.14)). Early-stage mergers may have a relatively flat distribution closer to the AGN line,  while late-stage mergers have a wider distribution in the $y$-axis, up to the bulges line. This is consistent with the sample signposting part of the transition between AGN and elliptical galaxies. It also may suggest a period of substantial black hole growth, starting close to nuclear coalescence.

\begin{figure*}
\begin{center}
\includegraphics[width=14cm]{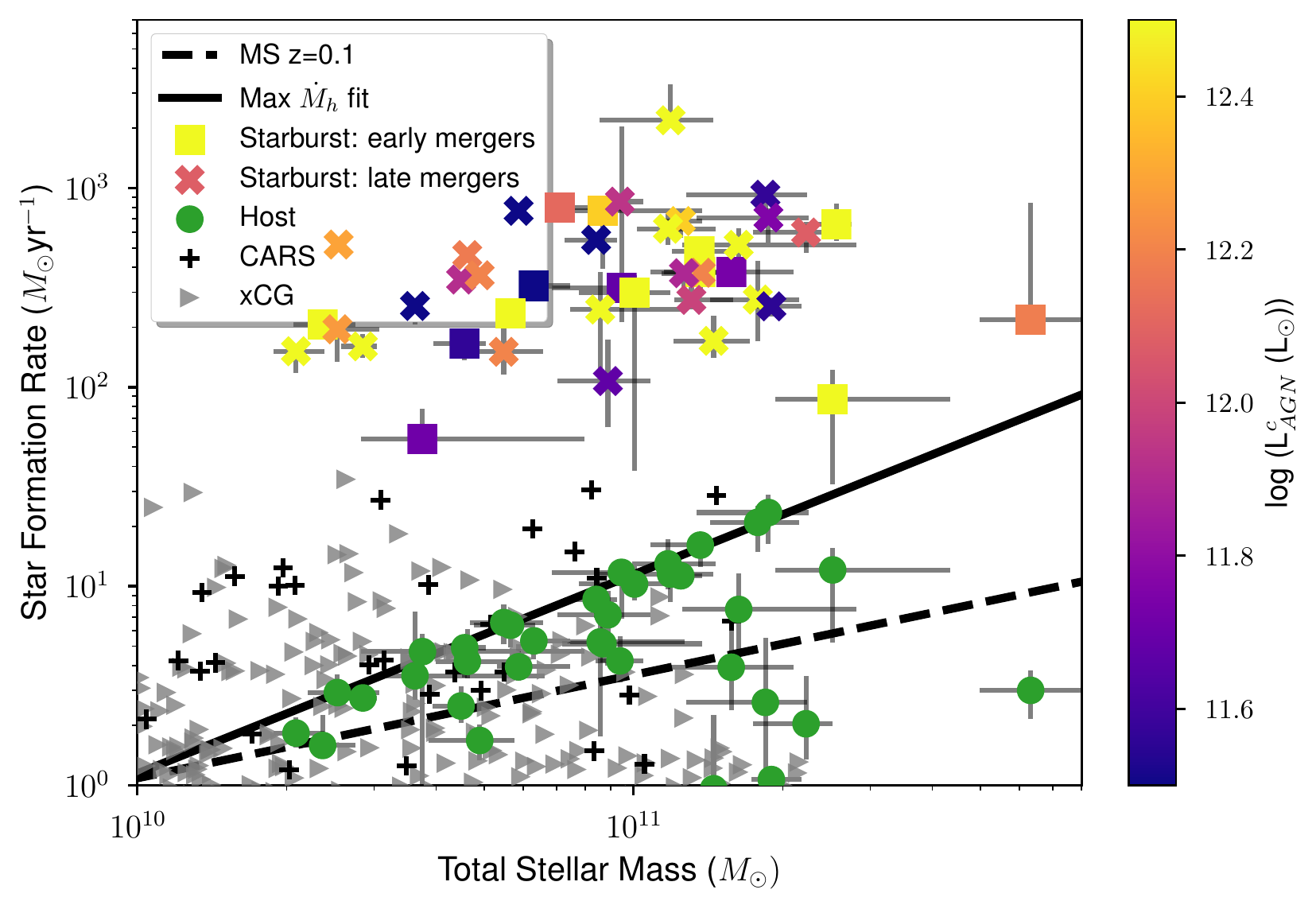}
\caption{The $\dot{M}_{*}$ - $M_{*}$ plane for our sample (\S\ref{sfmainseq}). Also plotted are the $\dot{M}_{*}$ - $M_{*}$ `main sequence' relations at $z=0$ and $z=0.1$, calculated via equation 28 of \citealt{spea14} with $t=13.70$\,Gyr and $t=12.36$\,Gyr, respectively. The host SFRs of our sample are enhanced by a factor $\sim 1.8$ compared to the $z=0.1$ main sequence, while starburst SFRs in the starburst are enhanced by a factor of $\sim 100$. Compared to the xCG and CARS samples, the host SFRs of our samples are fairly typical of the SFRs seen in these samples, but the starburst SFRs are in all cases higher. There is no apparent dependence on how far above the main sequence the starburst SFRs are on AGN luminosity. Neither is there a strong dependence of the starburst SFRs on merger stage.}
\label{fig:mstarmainseq}
\end{center}
\end{figure*}

\section{Star formation vs. AGN activity}\label{sec:starform}
We here examine the relations between luminous activity, and the mass the activity may be assembling. First, we study star formation in context with both total and starburst stellar mass. Second, we examine the relation between bolometric AGN luminosity and black hole mass.

\subsection{Star formation}\label{sfmainseq}
The sample have globally enhanced SFRs relative to the low-redshift $\dot{M}_{*}$ - $M_{*}$ main sequence (Figure \ref{fig:mstarmainseq}). Starburst SFRs are enhanced relative to the main sequence by about two orders of magnitude at $M_{*} = 10^{11}M_{\odot}$. There is no evidence for a relation between $M_{*}$ and starburst SFR. Host SFRs are enhanced by $\dot{M}_{*} \simeq 1.8\dot{M}_{MS}$, and are comparable to the higher (total) SFRs seen in the xCG and CARS samples. There may also be a stellar mass dependent `maximum' $\dot{M}_{*}$, corresponding to:

\begin{equation}
\dot{M}_{*,max} = 1.22 \pm 0.02 \times10^{-10} M_{*}
\end{equation}

\noindent in units of Solar masses (per year). This suggests a correlation between $M_{*}$ and $\dot{M}_{*}$, with a slope steeper than that of the low-redshift main sequence ($0.51\pm\sim0.2$ at $z=0.1$, using equation 28 of \citealt{spea14}).

There is no strong dependence of the starburst SFRs in Figure \ref{fig:mstarmainseq} on merger stage. This is consistent with star formation being triggered at any point in the merger, with perhaps a mild preference for more luminous starbursts in late-stage mergers. 

We also see no relation between AGN luminosity and offset from the main-sequence. This is in contrast with e.g. \citet{grim20}, who do find such a relation for X-ray selected AGN at $z\sim1$. The $2-10$\,KeV X-ray luminosities of the two samples reside in the same range, though the range in our sample is narrower. It may be the case that the narrower range in our sample makes the trend seen by \citet{grim20} impossible to discern. It is also possible that the trend does not exist in our sample, and that this difference arises via the higher comoving gas fraction at $z\sim1$ compared to locally, which leads to a tighter relation between star formation and AGN activity at $z\sim1$.

\begin{figure}
\begin{center}
\includegraphics[width=8cm]{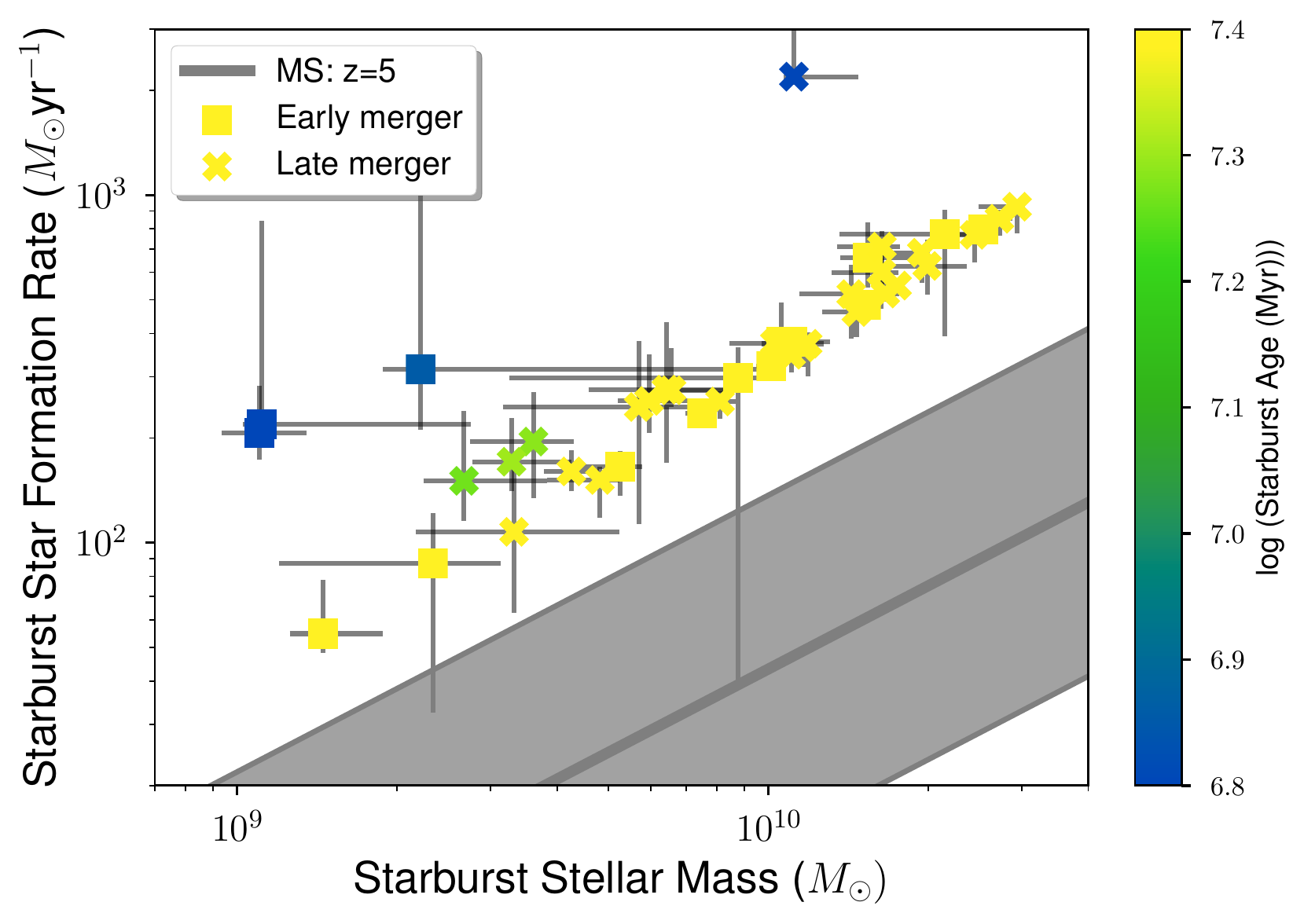}
\caption{The $\dot{M}_{*}$ - $M_{*,Sb}$ plane for our sample (\S\ref{sfmainseq}). Younger starbursts scatter significantly on this plane, but older starbursts settle onto a relation consistent with a stellar mass doubling time from the starburst of about $\sim10^{8}$ years. All of the sample lie above the star formation main sequence seen at $z=5$, consistent with there being a separate starburst mode at all epochs.}
\label{fig:mstarmainseq_starburst}
\end{center}
\end{figure}

We also consider the more focused comparison of the starburst star formation rate, plotted against the stellar mass in the starburst (Figure \ref{fig:mstarmainseq_starburst}). The youngest starbursts exhibit significant scatter, arising from the assumed exponential nature of the burst, but the older starbursts settle onto a tight relation. The slope of this relation implies a stellar mass doubling time of a little under $\sim10^{8}$ years, which indirectly infers, together with the ages of the starbursts (EFS21) that the starbursts in our sample are unlikely to increase the total stellar mass by more than about a factor of two. It is notable that the systems in this plot imply a consistent slope with the high-redshift main sequence, but still lie significantly above it. This is consistent with even high-redshift main-sequence star-forming galaxies not being ``global starbursts'', but instead having a distinct starburst and host star formation mode.

\begin{figure*}
\begin{center}
\includegraphics[width=14cm]{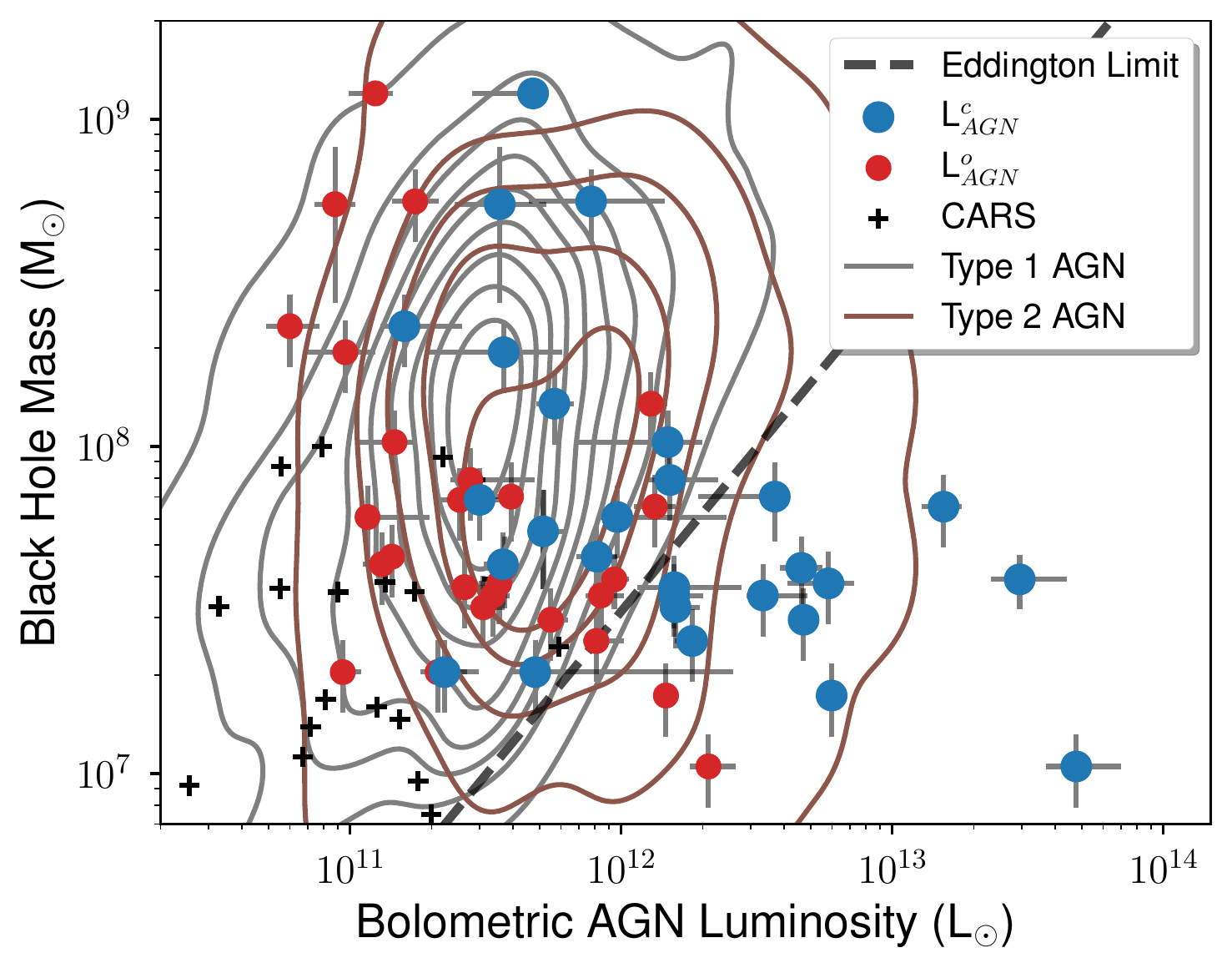}\\
\includegraphics[width=14cm]{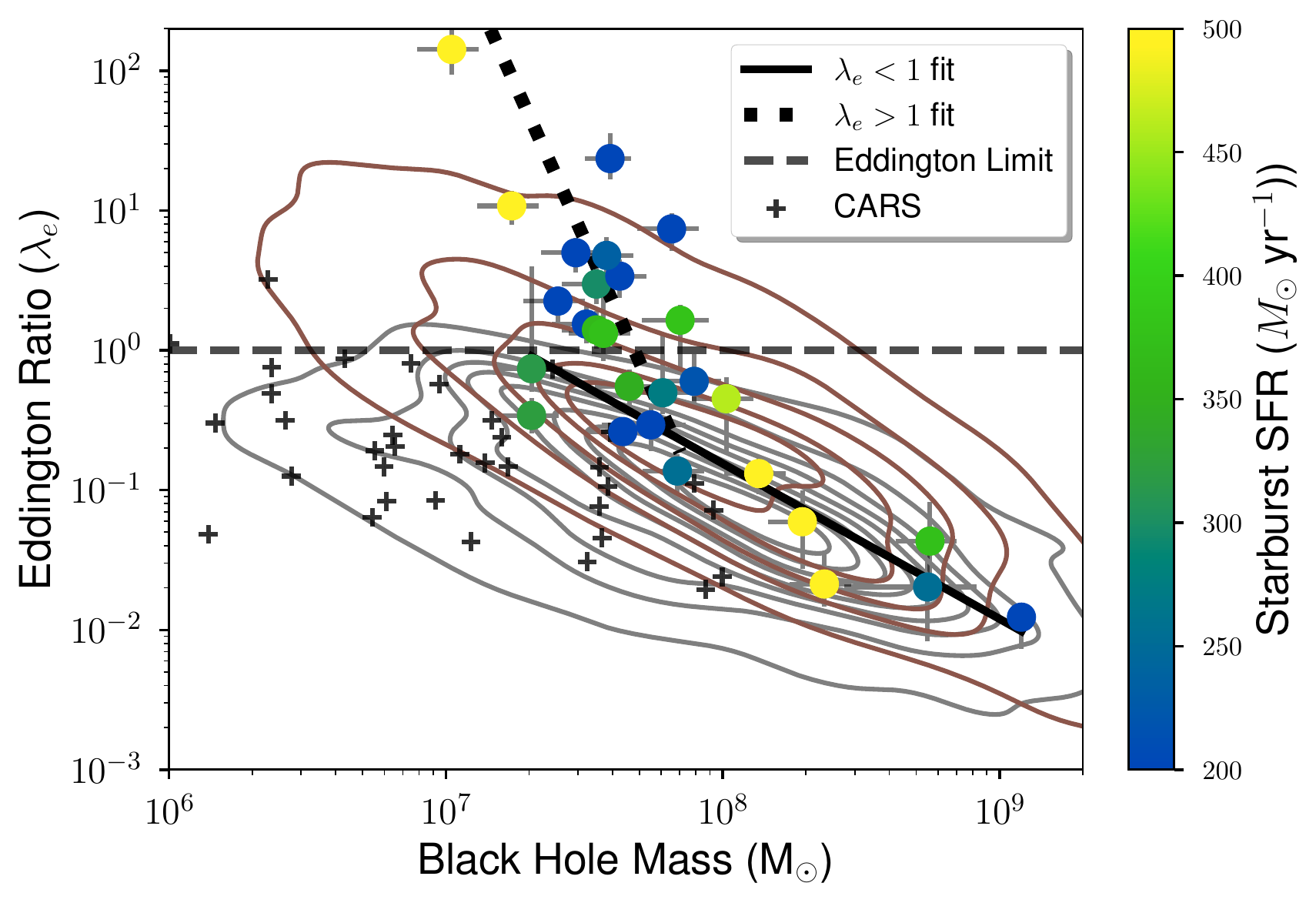}
\caption{Two slices through the $\rm{L}_{AGN}$ - $\rm{M}_{BH}$ - $\lambda_{e}$ plane (\S\ref{agnacti}), comparing our sample to the three AGN comparison samples (\S\ref{compsamp}). The type 1 AGN \citep{shen11} are plotted as grey contours, while the type 2 AGN \citep{kong18} are the brown contours. {\itshape Top:} Observed and anisotropy-corrected bolometric AGN luminosity vs. black hole mass. The observed AGN luminosities are consistent with those of the type 1 AGN, and higher, on average, than those of the CARS AGN. The corrected luminosities are more consistent with type 2 AGN, and infer super-Eddington accretion in 12 objects. {\itshape Bottom:} The $\lambda_{e} - \rm{M}_{BH}$ plane (using $\rm{L}^{c}_{AGN}$ to calculate $\lambda_{e}$ for our sample). The sub-Eddington objects coincide with the type 1 AGN, and can be characterized by the relation in Equation \ref{eq:subedd} (dashed line). The super-Eddington objects are offset from the type 1 AGN, and can be charactedized by the relation in \ref{eq:superedd} (solid line). The fits are plotted over the x-axis range of the data in each sub-sample, so they do not intersect at $\lambda_{e}=1$. The CARS AGN are consistent with being the lower black hole mass analogues of our sample. We observe no dependence of the positions of objects in this plane on starburst star formation rate.}
\label{fig:AGNbh}
\end{center}
\end{figure*}

\subsection{AGN activity}\label{agnacti}
We first consider the distribution of the sample in the $\rm{L}_{AGN}$ - $\rm{M}_{BH}$ plane (Figure \ref{fig:AGNbh}, top). The majority of the sample reside where the type 1 (grey contours, \citealt{shen11}) and type 2 AGN (brown contours, \citealt{kong18}) comparison samples are located. Our sample is consistent with harbouring more massive black holes powering higher luminosity AGN, than the CARS sample. The Eddington ratios of our sample are however sometimes very high. Three objects exhibit super-Eddington accretion when using $\rm{L}^{o}_{AGN}$, rising to twelve when using $\rm{L}^{c}_{AGN}$. 

At face value the sample show a mild anti-correlation between $\rm{L}_{AGN}$ and $\rm{M}_{BH}$ ($\tau$, $p$) = (-0.33, 0.01) for L$^{o}_{AGN}$ and ($\tau$, $p$) = (-0.36, 0.01) for L$^{c}_{AGN}$). However, the super- and sub-Eddington objects may have differing distributions in the $\rm{L}_{AGN}$ - $\rm{M}_{BH}$ plane. The super-Eddington objects are mostly distributed where the type 2 comparison AGN reside, with a few objects at even higher luminosities, whereas the sub-Eddington objects are more similar to the type 1 quasars. Neither sub-sample shows a statistically significant relation between $\rm{L}_{AGN}$ and $\rm{M}_{BH}$ (e.g. for L$^{c}_{AGN}$: ) ($\tau$, $p$) = (-0.24, 0.56) and (-0.15, 0.33) for the super- and sub-Eddington samples, respectively\footnote{Instead dividing the sample in two at L$^{c}_{AGN}\simeq2\times10^{12}$L$_{\odot}$ or $\rm{M}_{BH}\simeq10^{8}$M$_{\odot}$ gives a similar result.}). We conclude that the apparent anti-correlation in the whole sample arises from these two sub-samples being considered together. 

We plot Eddington ratio against black hole mass in Figure \ref{fig:AGNbh}, bottom. Adopting a log-linear relation as a phenomenological model, then there is evidence for different slopes in the $\lambda_{e} - \rm{M}_{BH}$ plane among the super- and sub-Eddington systems, with:

\begin{equation}\label{eq:subedd}
\log{\lambda_{e}} = (-1.10\pm0.14) \log{\left(\frac{\rm{M}_{BH}}{\rm{M}_{\odot}}\right)} + 8.07\pm1.09
\end{equation}

\noindent for the sub-Eddington systems (dashed line) and:

\begin{equation}\label{eq:superedd}
\log{\lambda_{e}} = (-3.73 \pm 1.16)  \log{\left(\frac{\rm{M}_{BH}}{\rm{M}_{\odot}}\right)} + 28.92\pm8.69
\end{equation}

\noindent for the super-Eddington systems (solid line). 

The sub-Eddington objects are approximately co-located with the type 1 and type 2 AGN comparison samples, while the super-Eddington objects overlap with some of the type 2 AGN. Our sample is consistent with being the higher black hole mass analogue of the CARS AGN sample. There is however no clear delineation between the super-Eddington systems and sub-Eddington systems on star formation rate (though see e.g. \citealt{tor21}).

\section{Mass assembly across the merger sequence}\label{massassemb}
We here examine the pattern of stellar and black hole mass assembly across the merger sequence. To obtain a finer estimate of merger stage, we here use projected nuclear separation as a proxy for merger stage, though the results are consistent if using the MS values instead. 

\subsection{Stellar mass assembly}
\begin{figure*}
\begin{center}
\includegraphics[width=8cm]{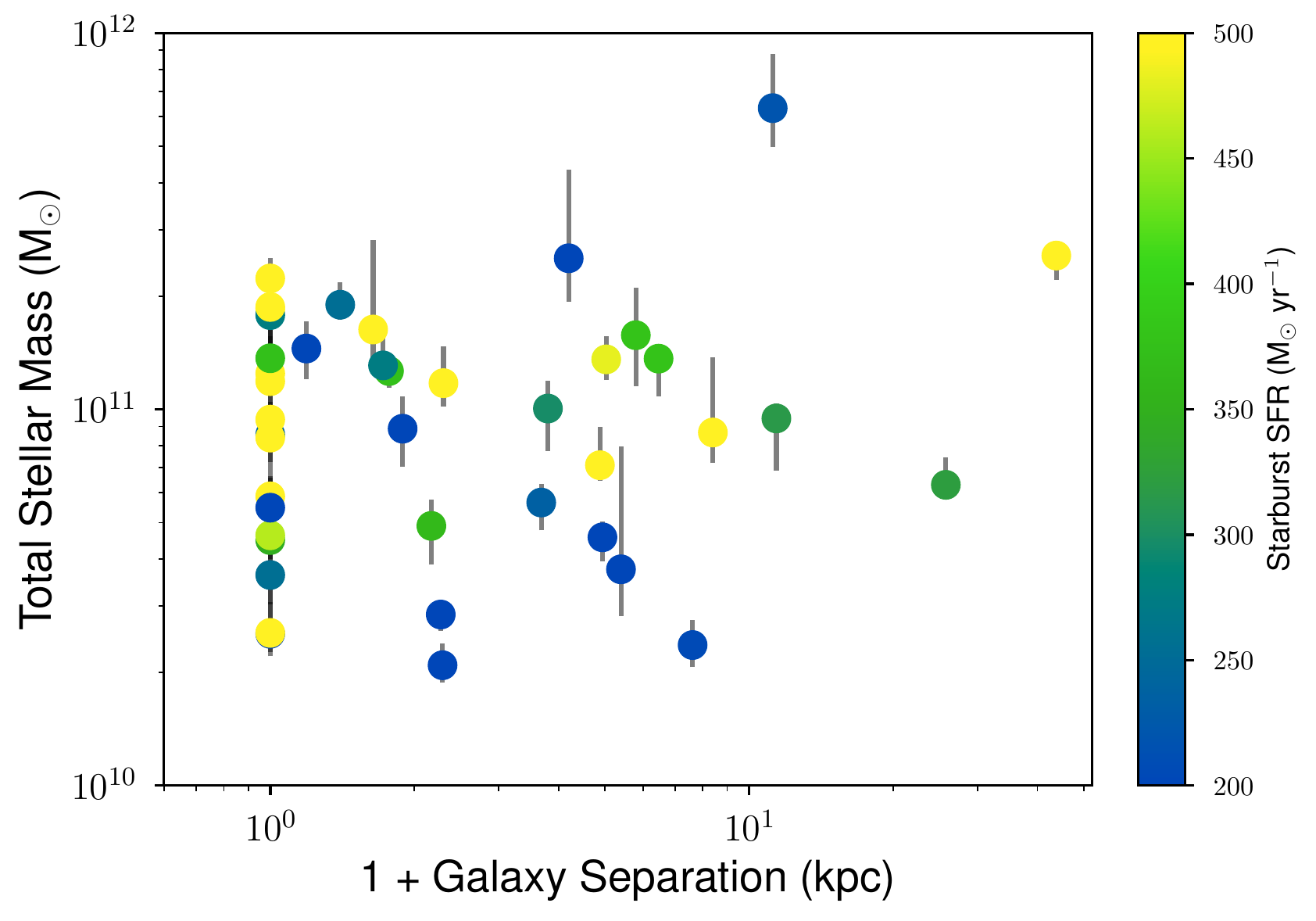} 
\includegraphics[width=8cm]{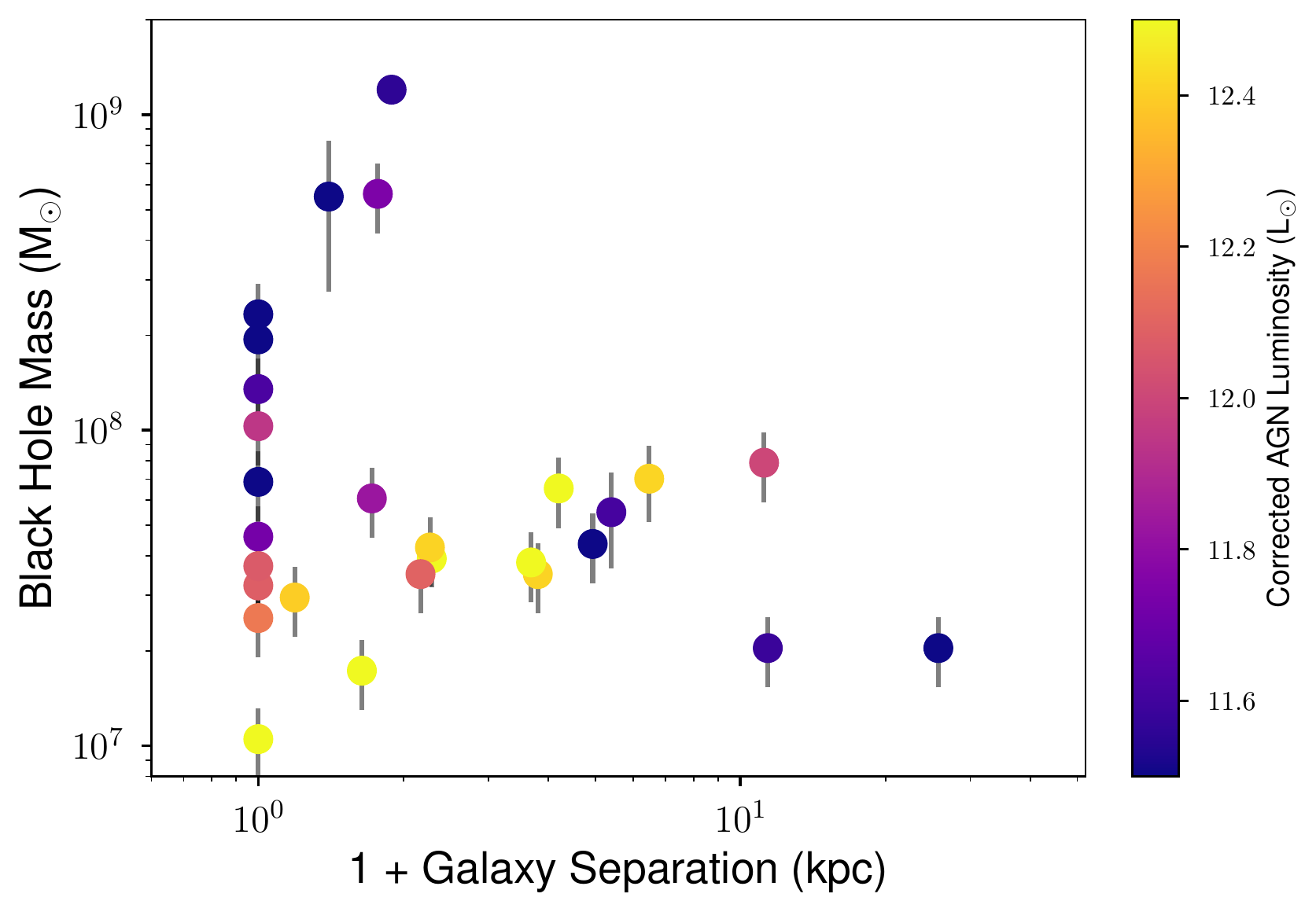} \\
\includegraphics[width=8cm]{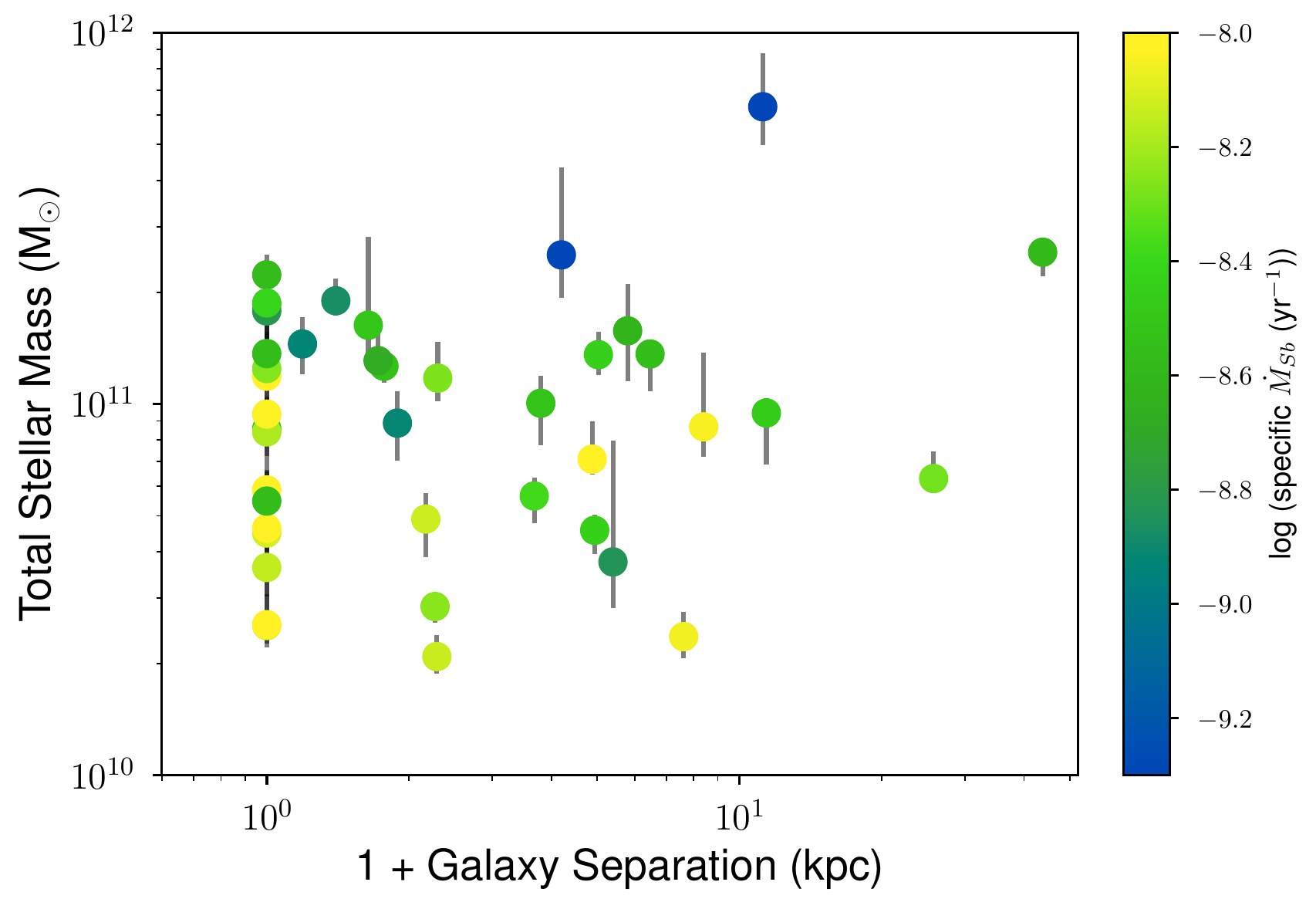} 
\includegraphics[width=8cm]{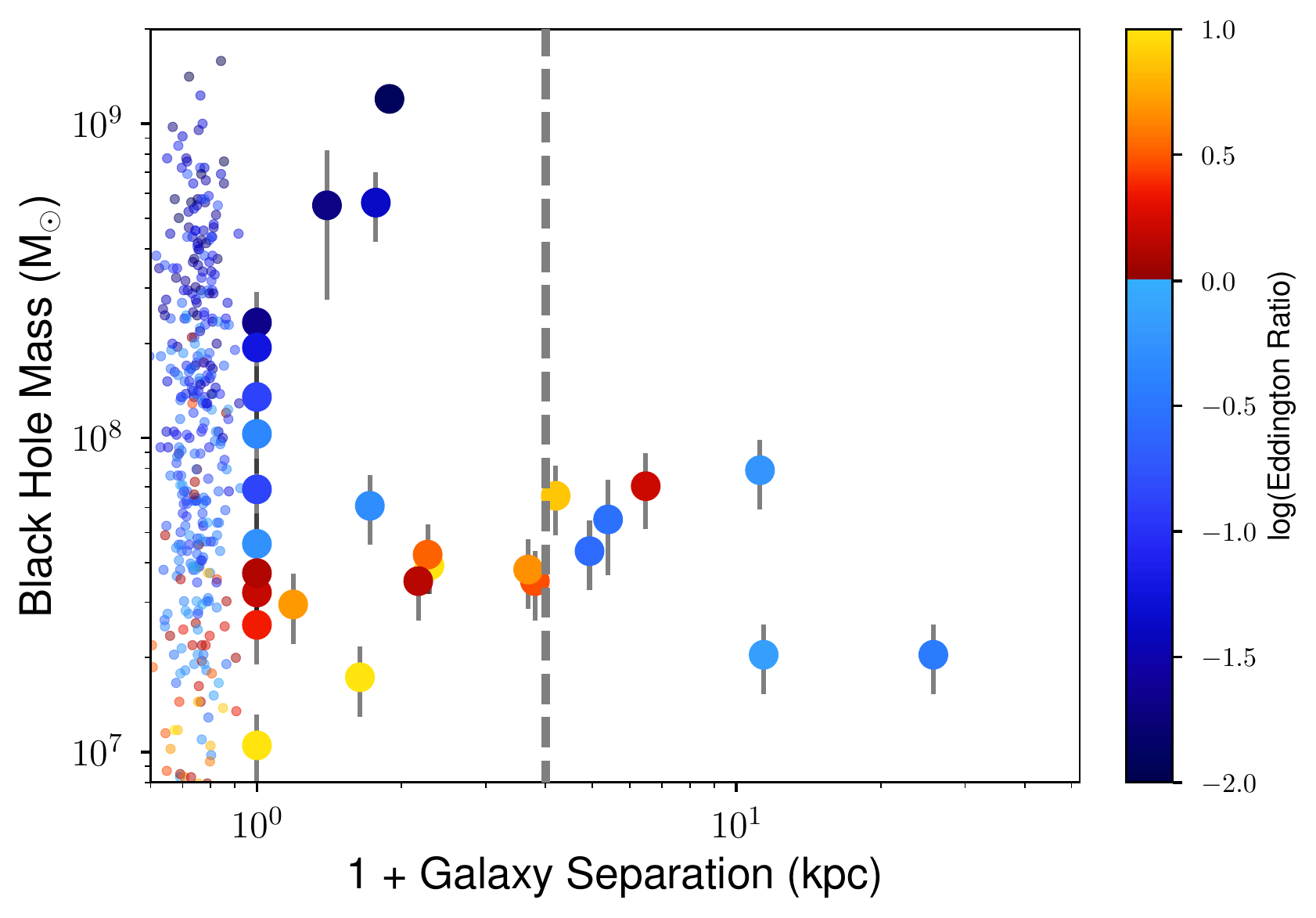} \\
\caption{{\itshape Left} stellar and {\itshape right} black hole mass as a function of nuclear separation (\S\ref{massassemb}). The small points in the right-hand column all at $(1+n_{s})\simeq10^{0}$\,kpc are the type 2 AGN comparison sample  with artificial nuclear separations to allow comparisons with the Eddington ratios as a function of black hole mass. There is mild evidence that luminous starbursts prefer later-stage mergers, but no strong evidence for substantial stellar mass assembly from star formation. Conversely, super-Eddington accretion is associated with late-stage mergers and $\rm{M}_{BH}<10^{8}$\,$\rm{M}_{\odot}$. The sub-Eddington systems have a wider spread in the $\rm{M}_{BH} - n_{s}$ plane. The vertical dashed line indicates the boundary in nuclear separation used for the calculations in \S\ref{macccalc} - \ref{acceffcalc}.}
\label{fig:merge_agn1}
\end{center}
\end{figure*}

The left column of Figure \ref{fig:merge_agn1} presents nuclear separation against stellar mass, coded by SFR and specific SFR in the starburst. Higher SFRs show a mild preference for smaller nuclear separations, consistent with the results in \S\ref{sfmainseq}, and with some previous studies \citep[e.g.]{loo82,alon16}. It is also consistent with the merger itself not acting as a quenching mechanism (see also e.g. \citealt{weig17}). We also observe no trends in starburst age with nuclear separation. 

These results are consistent with star formation not generating substantial new stellar mass during the merger, relative to what is present in the progenitors. Including a potential stellar mass increase from double- to single-nucleus systems of a factor $1.5 - 2$ from the merger itself, then we infer a soft upper limit on the stellar mass increase due to star formation of about $20\%$. This limit is consistent with the typical starburst SFRs and ages of our sample (EFS21). This result does however contrast with mergers at only slightly higher redshifts, which may assemble significant new stellar mass from star formation \citep{cana13}.

There is another plausible explanation for the results in the left column of Figure \ref{fig:merge_agn1}. Starbursts may assemble significant new stellar mass in events triggered throughout the merger, on timescales shorter than the merger itself. Such a pattern of activity might not show a clear rise in stellar mass as a function of nuclear separation. We consider this explanation less likely though as the starburst ages (EFS21) are within about an order of magnitude of the merger timescales from simulations.

\subsection{Black hole mass assembly}\label{bhmass}
In the right column of Figure \ref{fig:merge_agn1} we plot nuclear separation against black hole mass, coded by $L^{c}_{AGN}$ and Eddington ratio. The small points all at $(1+n_{s})\simeq10^{0}$\,kpc are the type 2 AGN comparison sample with artificial nuclear separations to allow comparisons with the Eddington ratios as a function of black hole mass. Higher AGN luminosities prefer both smaller nuclear separations and less massive black holes. Super-Eddington AGN reside in mergers with nuclear separations smaller than $\sim8$\,kpc, suggesting that the super-Eddingon phase commences close to galactic coalescence.  

There are several potential biases that may affect our results, and alternative explanations to the ones we present here. We summarize these in the appendix. They include; the choice of AGN radiative transfer model (\S\ref{appradtrans}), the choice of black hole mass estimates (\S\ref{appbhmass}), systematics in black hole mass estimates (\S\ref{appsystem}), black hole mergers (\S\ref{appmergers}), sample bias (\S\ref{abias}), the implications of the absence of  super-Eddington type 1 quasars (\S\ref{appsequso}), inclination angle bias (\S\ref{appinclbias}), transient phenomena (\S\ref{apptransi}), host obscuration degeneracy (\S\ref{apphostobs}), and deviation from Thomson Scattering (\S\ref{appthomdev}).

From our results, it is possible to estimate the duration, mass accreted, and efficiency of the super-Eddington phase. These estimates are crude and have substantial uncertainty, but they are currently the best observational constraints for low-redshift infrared-luminous mergers, so we present them in the following. 

\subsubsection{Mass accreted}\label{macccalc}
The maximum nuclear separation of a super-Eddington system in Figure \ref{fig:merge_agn1} is $\sim$8\,kpc, but the individual Eddington ratios are uncertain, and nuclear separations are hard to measure even in high resolution data. For consistency with the boundary in nuclear separation between merger stages, we adopt a maximum separation of a $\lambda_{e}>1$ system to be $3$\,kpc. 

We estimate the average black hole mass before and after the super-Eddington accretion phase as follows. Systems yet to pass through a super-Eddington phase will have $n_{s}>3$\,kpc and $\lambda_{e}<1$, for which we find:

\begin{equation}\label{eq:massbefore}
\langle \rm{M}_{BH}(before) \rangle = 4.37\pm1.11\times10^{7}\rm{M}_{\odot}
\end{equation}

\noindent Conversely, post super-Eddington systems will have a single nucleus and $\lambda_{e}<1$. Additionally, simulations suggest that an SMBH-SMBH merger can take place on the required timescale \citep{sob21} so we assume that it does and that it contributes a 25\% mass increase at the end of the super-Eddington phase (allowing for unequal mass mergers). The mass at the end of the super-Eddington phase is then:

\begin{equation}\label{eq:massafter}
\langle \rm{M}_{BH}(after) \rangle = 9.75\pm2.22\times10^{7}\rm{M}_{\odot}
\end{equation}

\noindent The mass accreted during the super-Eddington phase is thus:

\begin{equation}\label{eq:massacc}
\langle \Delta \rm{M}_{BH}\rangle = \langle \rm{M}_{BH}(after) \rangle - \langle \rm{M}_{BH}(before) \rangle = 5.38\pm2.48\times10^{7}\rm{M}_{\odot}
\end{equation}

\noindent Varying the adopted boundary in nuclear separation between the super- and sub-Eddington phases over $2-6$\,kpc causes $\langle \Delta \rm{M}_{BH}\rangle$ to vary by about 15\%.

\subsubsection{Duration}\label{sec:durat}
To estimate the duration of the super-Eddington phase we employ results from galaxy merger simulations \citep{deb11}. These suggest that the infrared-luminous, $<3$kpc nuclear separation phase of the merger lasts approximately 80Myr. 

From Table \ref{tablethesample}, 21 objects have both a measured black hole mass and a nuclear separations of $<3$kpc. Of these, nine exhibit super-Eddington accretion rates. A crude estimate of the duration of the super-Eddington phase is thus $t_{e} = 44$\,Myr. We adopt an uncertainty on this value of 50\%.

\subsubsection{Accretion efficiency}\label{acceffcalc}
We first estimate the mean accretion rate during the super-Eddington phase by combining Equation \ref{eq:massacc} with the estimate of the duration of the super-Eddington phase in \S\ref{sec:durat}. This yields:

\begin{equation}\label{eq:massaccrate}
\langle \rm{\dot{M}}_{BH,SE}\rangle = \frac
{\langle \Delta \rm{M}_{BH}\rangle}{t_{e}} = 1.57\pm1.07 M_{\odot}yr^{-1}
 \end{equation}

\noindent We then calculate the mean bolometric AGN luminosity of super-Eddington objects with $n_{s}<3$: 

\begin{equation}\label{eq:meanlse}
\langle \rm{L}^{c}_{AGN}(n_{s}<3, \lambda_{e}>1)\rangle = 9.81\pm3.83\times10^{12}\,L_{\odot}
\end{equation}

\noindent which we assume is not saturated due to photon trapping in the accretion flow. We then estimate the accretion efficiency $\epsilon$ - the ratio of the energy radiated per unit time to the energy that would be radiated assuming mass-energy equivalence - as:

\begin{equation}\label{eq:acceff}
\epsilon = \frac
{\langle \rm{L}^{c}_{AGN}(n_{s}<3, \lambda_{e}>1)\rangle}
{\langle \rm{\dot{M}}_{BH,SE}\rangle c^{2}} = (42\pm33)\%
\end{equation}

\section{A hidden population of super-Eddington AGN?}\label{hidpop}
We identify a significant fraction of late-stage infrared-luminous mergers as harboring super-Eddington accretion. This result is consistent with results from several recent models:

\begin{itemize}

\item Super-Eddington growth is more common for less massive black holes \citep{shira19}.

\item The super-Eddington phase occurs towards the end of the merger and grows SMBHs to masses of $10^{8}$\,$\rm{M}_{\odot}$ \citep{lili12}. 

\item The super-Eddington phase lasts up to $\sim$\,10\,Myr (\citealt{volon15}, though we may prefer a somewhat longer duration). 

\item Super-Eddington AGN have only weak observed emission in the X-ray through optical, but are more luminous at longer wavelengths (\citealt{pog20}, see also \citealt{teng15,laha18,yamada21}). 

\item There are simulations that find rapid, self-regulated growth of SMBHs with masses $\lesssim10^{8.5}$M$_{\odot}$ in mergers \citep{wein18}. 

\end{itemize}

The accretion efficiency we derive is high, and implies remarkably efficient accretion during the super-Eddington phase. It is however consistent with the expected limit for a maximally rotating Kerr black hole \citep{bar72}. It is higher than, though consistent with, the accretion efficiency inferred from cosmological constraints from the SMBH mass density and the integrated luminosity from AGN \citep{sol82}\footnote{It has recently been proposed that all black holes cosmologically couple and thus gain mass as the cosmological scale factor increases \citep{cro19,cro20,cro21}. This cosmological mass increase will cause Keplerian orbits around SMBHs to adiabatically collapse. The duration of the super-Eddington phase, however, is short compared to changes in the scale factor at $z < 0.3$. We therefore do not consider cosmological coupling here.}.

Our result increases the number of candidate super-Eddington systems; other candidates include quasars at $z\gtrsim6$ \citep{tang19,yang21}, some type 2 AGN \citep{kawa04,kong18}, and some X-ray selected AGN \citep{coff19}. The Eddington ratios we derive are also in line with these studies, and are not even the most extreme observed. For example, the AGN IRAS 04416+1215 has an inferred Eddington ratio in excess of 400 \citep{torto22}. 

Our result is also implicitly consistent with several observational studies. First, collapsing our sample along the x-axis of Figure \ref{fig:merge_agn1} bottom right yields a distribution that is consistent with type 2 AGN. Second, three of our sample (\# 6, 37, 38) have $\lambda_{e}$ estimates from hard X-ray observations, which are independent of our method. All three are consistent within $2\sigma$ of those we derive (\citealt{teng15}; 1.2, 5.2, 0.04, respectively). Finally, the range in projected nuclear separations over which we observe super-Eddington accretion is consistent with the range over which an increased fraction of Compton-thick AGN is observed in a sample that overlaps ours \citep{ricci21}.

Some broader implications are apparent from this result. Infrared-luminous galaxies are rare at $z<0.3$ but become much more common by $z\gtrsim1.0$. While the merger fraction at high redshift is uncertain (e.g. \citealt{ryan08,stott13,beth15,man16}), our discovery is consistent with a super-Eddington phase in a significant fraction of these higher redshift systems, and that super-Eddington accretion is a significant mode for the cosmological growth of the SMBH mass function. This may require revision of galaxy evolution models which cap SMBH accretion at near the Eddington limit \citep{som08,shan13,dub14,hen15,scha15}.  It also provides validation for the idea of super-Eddington accretion to explain the existence of SMBHs in quasars at $z\gtrsim5$, without recourse to direct collapse models. It also provides a physical explanation for the $\rm{M}_{BH}$ vs. $\rm{L}_{IR}/\rm{L}_{e}$ anticorrelation \citep{kawa07}. It is consistent with the bulk of cosmological SMBH growth being obscured \citep{mart05}, with more luminous AGN being found in later-stage mergers \citep{ricc17,koss18}, and with dusty quasars having high Eddington ratios \citep{urr12,kim15}.

\section{Conclusions} \label{sec:conclusions}
We have studied the assembly of stellar and black hole mass in low-redshift ultraluminous infrared galaxies, by combining radiative transfer modelling results with archival measures of molecular gas mass and central black hole mass. Our conclusions are:

1 - The $\rm{M}_{*} - \rm{M}_{H_{2}}$ distribution of our sample is consistent with massive, gas-rich galaxies at low redshift. More luminous starbursts show a mild preference for higher gas fractions. There is a mild correlation between $\rm{M}_{*}$ and $\rm{M}_{H_{2}}$, but the correlation does not strongly depend on merger stage. 

2 - The $M_{*} - M_{BH}$ distribution shows no evidence for a correlation. The distribution of late- vs. early-stage mergers is consistent with a transition phase between AGN and ellipticals, accompanied by substantial black hole mass growth. 

3 - The starburst SFRs in our sample are strongly enhanced relative to SFRs in the low-redshift main-sequence, by a factor of $\dot{M}_{Sb} \simeq 100\dot{M}_{MS}|_{10^{11}M_{*}}$. There is no evidence for a relation between $M_{*}$ and $\dot{M}_{Sb}$. Comparisons between starburst SFR, starburst age, and stellar mass assembled in the starburst suggest that the starburst increases the total stellar mass by a factor of about two or less over the course of the merger. 

4 - SFRs in the host galaxy are also enhanced relative to the low-redshift main-sequence, by $\dot{M}_{*} \simeq 1.8\dot{M}_{MS}$. The host galaxy SFRs are consistent with those in gas-rich galaxies at low redshift, and may exhibit a stellar mass dependent `maximum' $\dot{M}_{*}$. These results show no strong dependence on merger stage. 

5 - Directly examining star formation across the merger sequence shows that more luminous starbursts show a mild preference for advanced mergers, though they are found in all merger stages. There is however no strong evidence for a substantial increase in stellar mass due to star formation, though this does depend on the starburst duration compared to the merger duration. It is also consistent with the merger itself not acting as a quenching mechanism.

6 - Twelve of the sample show evidence for super-Eddington accretion. The sub-Eddington and super-Eddington samples may have different slopes in the $\lambda_{e} - \rm{M}_{BH}$ plane, suggesting that the super- and sub-Eddington accretion phases are distinct from each other. 

7 - Super-Eddington accretion rates may prefer smaller black holes in advanced mergers. We estimate the mean mass accreted during the super-Eddington phase to be $5.38\pm2.48\times10^{7}\rm{M}_{\odot}$, the duration of the super-Eddington phase to be $44\pm22$\,Myr, and the accretion efficiency during the super-Eddington phase to be $(42\pm33)\,\%$.

8 - Our results are consistent with super-Eddington accretion bringing the remnants of infrared-luminous mergers in line with the $M_{*} - M_{BH}$ relation in local ellipticals. This also suggests that super-Eddington accretion may be an important growth mode for black holes at all redshifts.

\section*{ACKNOWLEDGEMENTS}
We thank the referee for an extremely helpful report, and Joshua Barnes for several helpful discussions. AE acknowledges support from the project EXCELLENCE/1216/0207/ GRATOS funded by the Cyprus Research \& Innovation Foundation and the project CYGNUS funded by the European Space Agency. This work is based in part on observations made with the Spitzer Space Telescope, and the Herschel Space Observatory. Spitzer was operated by the Jet Propulsion Laboratory, California Institute of Technology under a contract with NASA. The Herschel spacecraft was designed, built, tested, and launched under a contract to ESA managed by the Herschel/Planck Project team by an industrial consortium under the overall responsibility of the prime contractor Thales Alenia Space (Cannes), and including Astrium (Friedrichshafen) responsible for the payload module and for system testing at spacecraft level, Thales Alenia Space (Turin) responsible for the service module, and Astrium (Toulouse) responsible for the telescope, with in excess of a hundred subcontractors.

\section*{Data Availability}
The data underlying this article are available in the article, and within the articles and their online supplementary materials cited in \S\ref{sec:methods}

\bibliographystyle{mnras}
\bibliography{Assembly_ulirgs_rev}  

\appendix

\begin{figure*}
\begin{center}
\includegraphics[width=7cm]{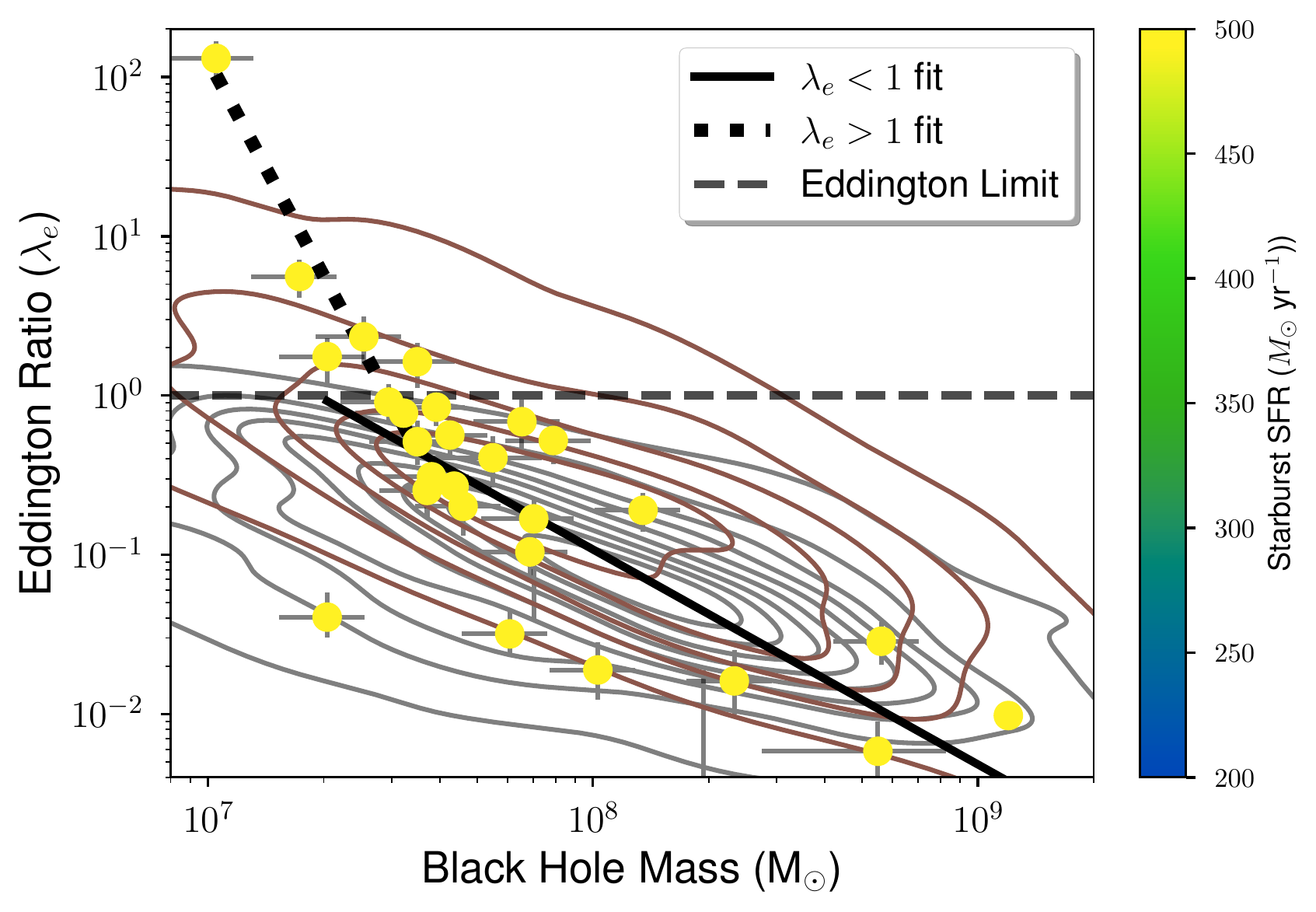}
\includegraphics[width=7cm]{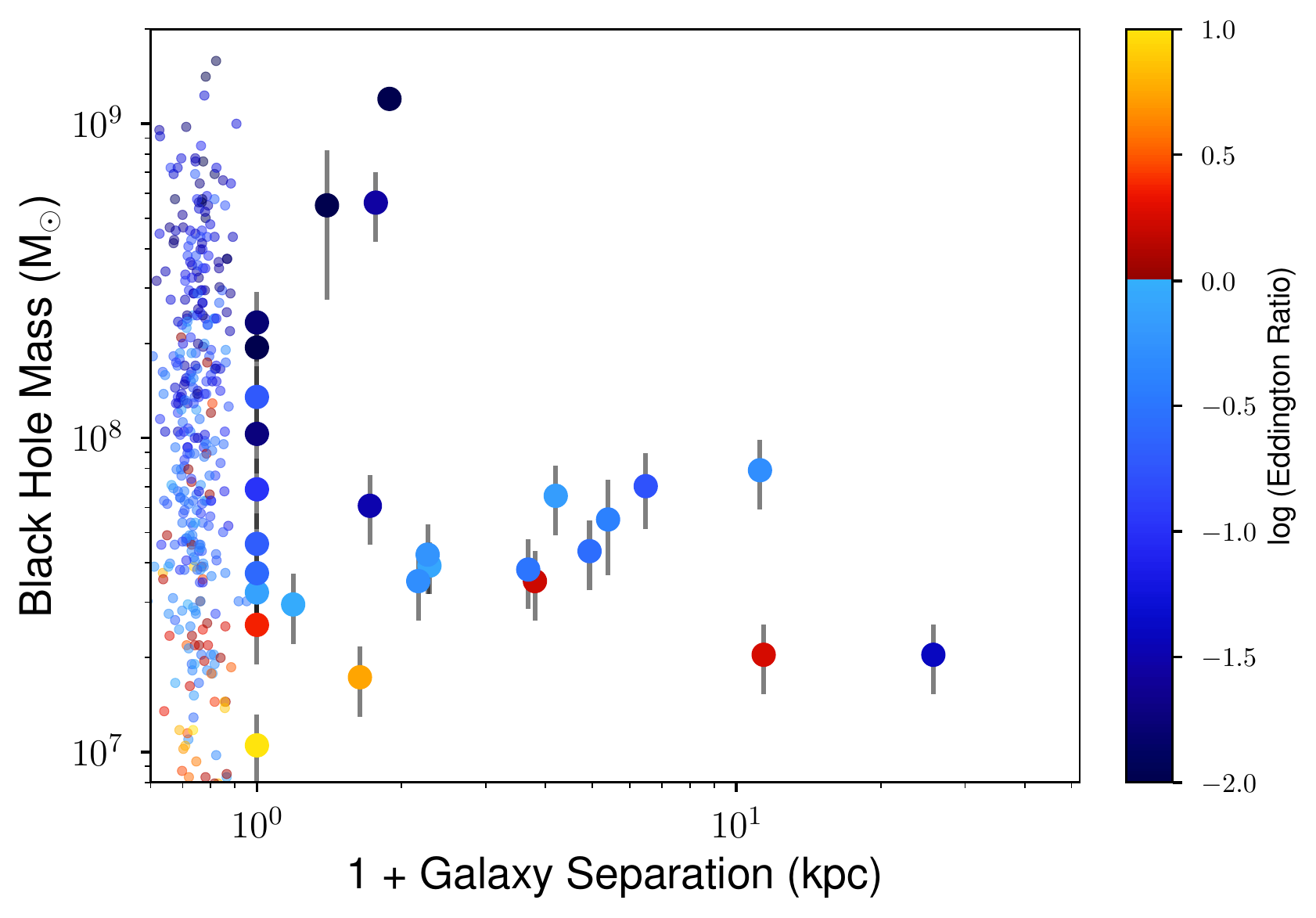}\\
\caption{Figures \ref{fig:AGNbh} lower and \ref{fig:merge_agn1}, made using fits that replace the CYGNUS AGN models with the \citealt{fritz06} AGN models. The results are effectively unchanged.}
\label{fig:altexfri}
\end{center}
\end{figure*}

\begin{figure*}
\begin{center}
\includegraphics[width=7cm]{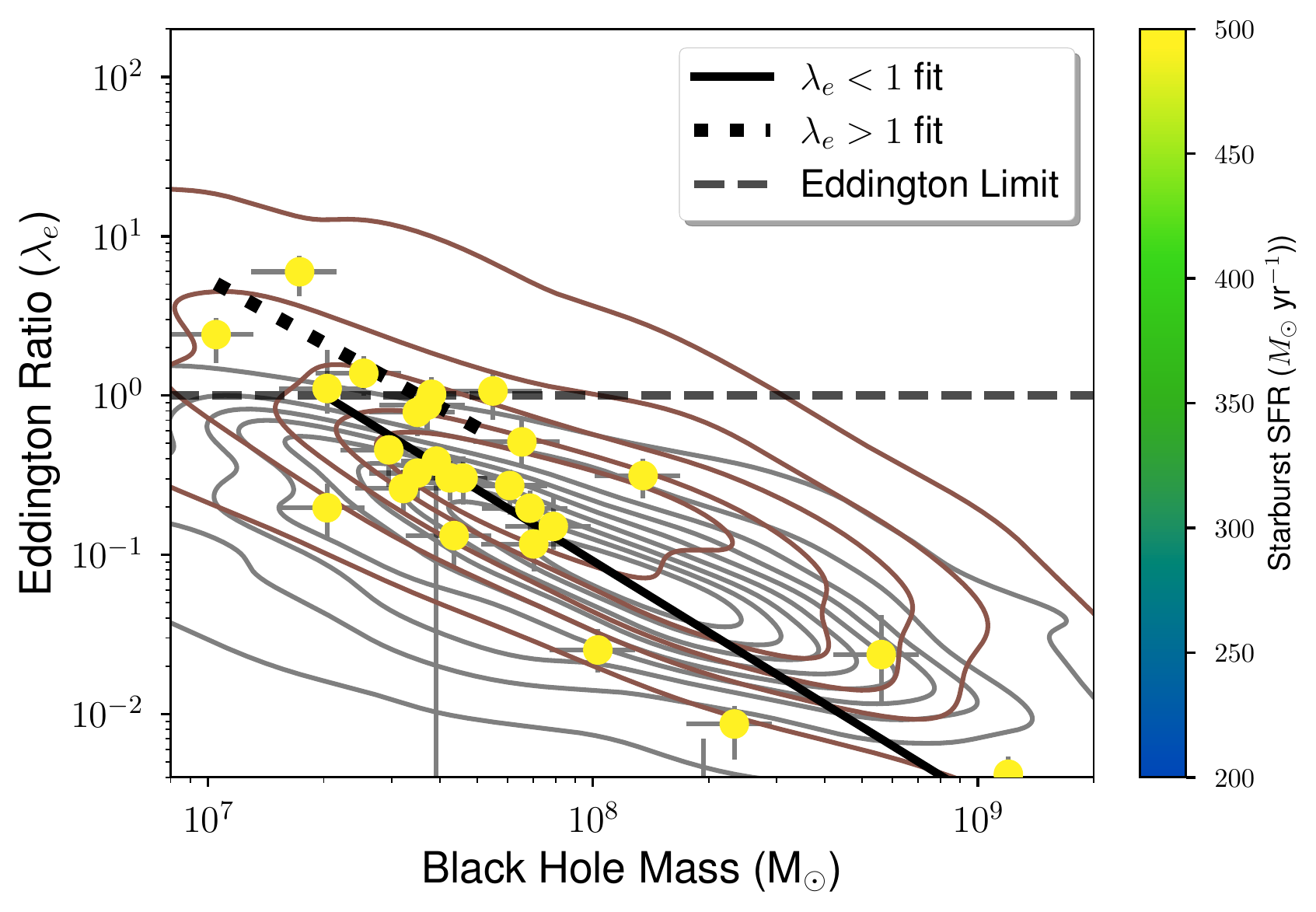}
\includegraphics[width=7cm]{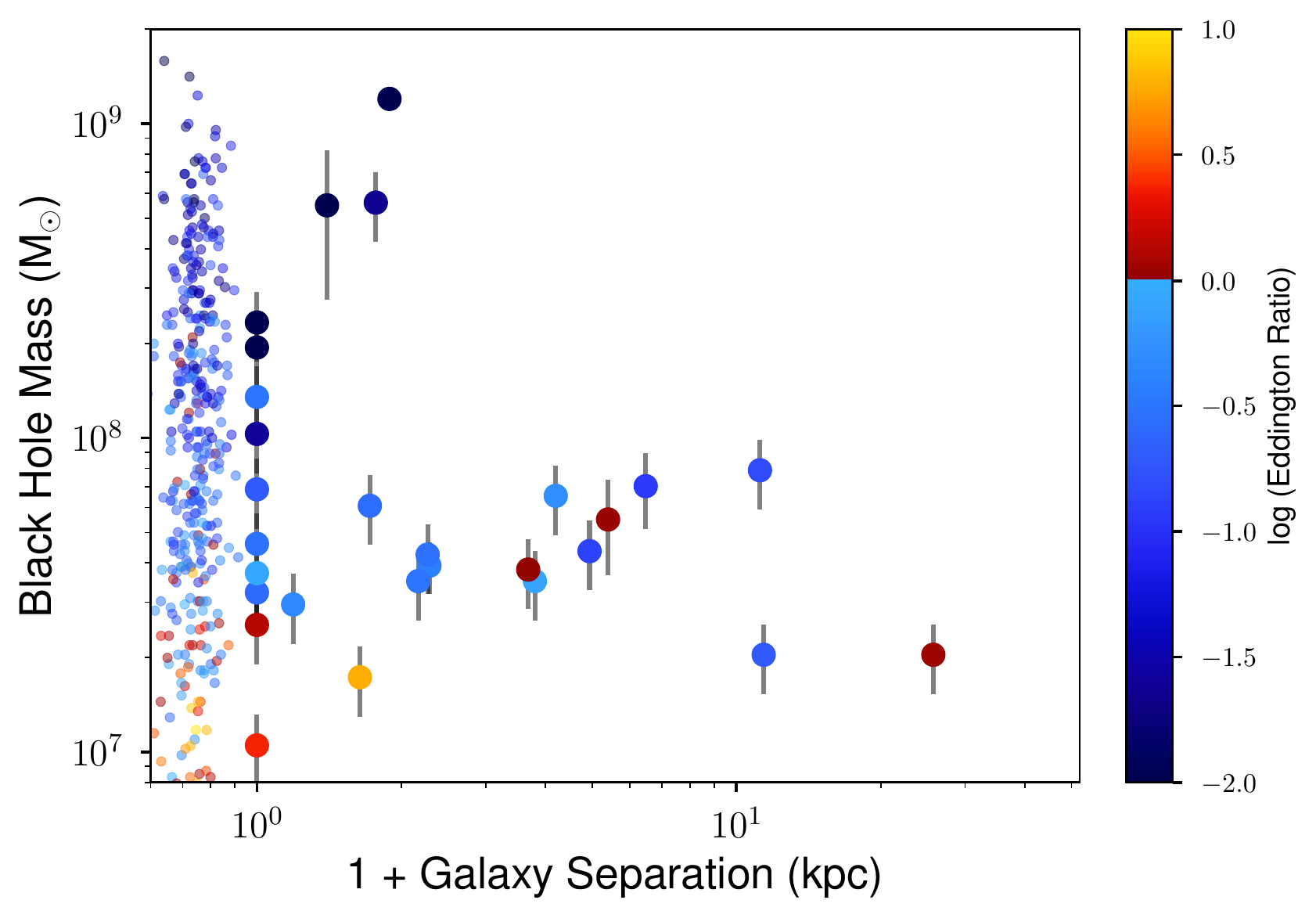}\\
\caption{Figures \ref{fig:AGNbh} lower and \ref{fig:merge_agn1}, made using fits that replace the CYGNUS AGN models with the \citealt{sieben15} two-phase AGN models.}
\label{fig:altexsie}
\end{center}
\end{figure*}

\begin{figure*}
\begin{center}
\includegraphics[width=7cm]{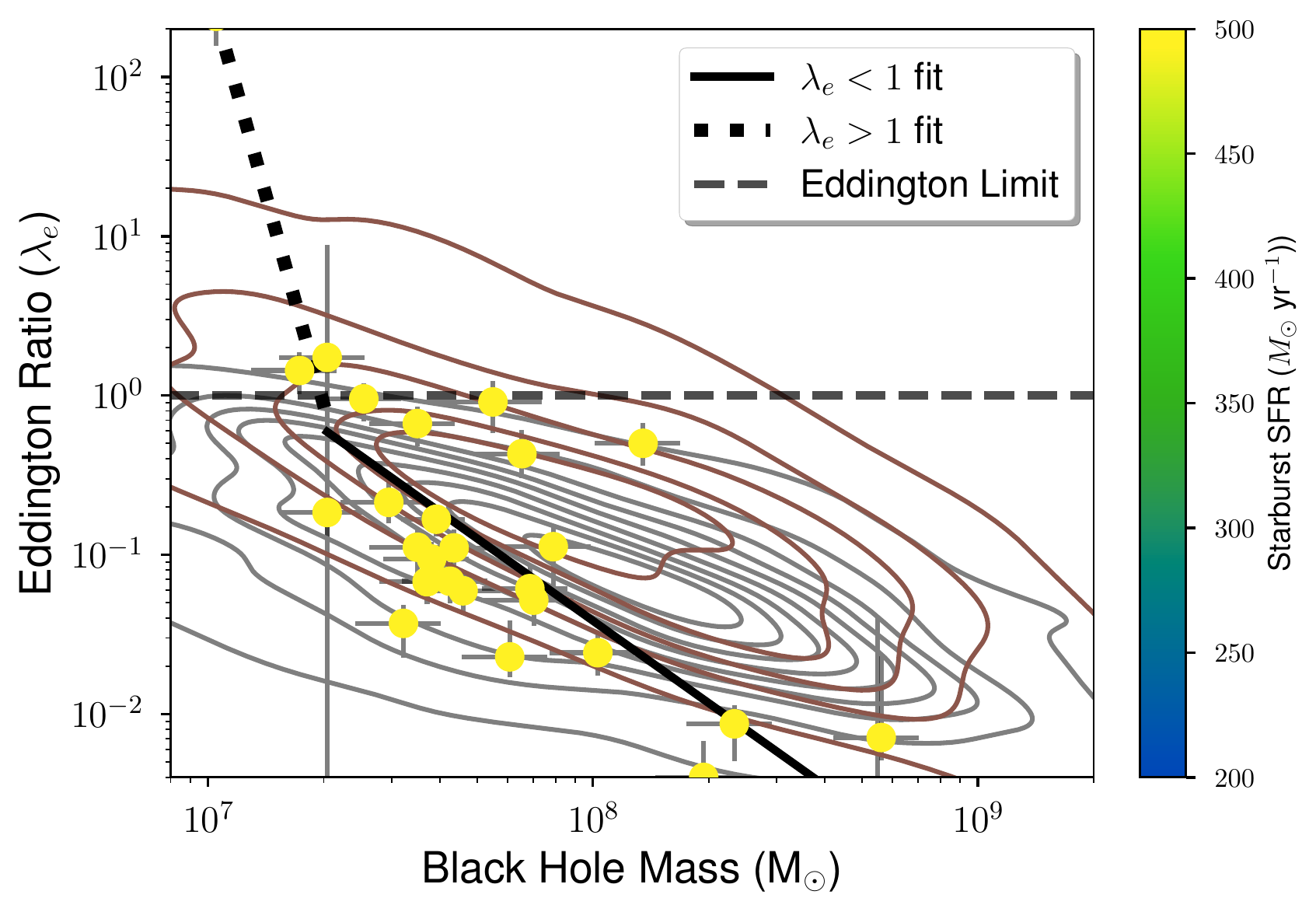}
\includegraphics[width=7cm]{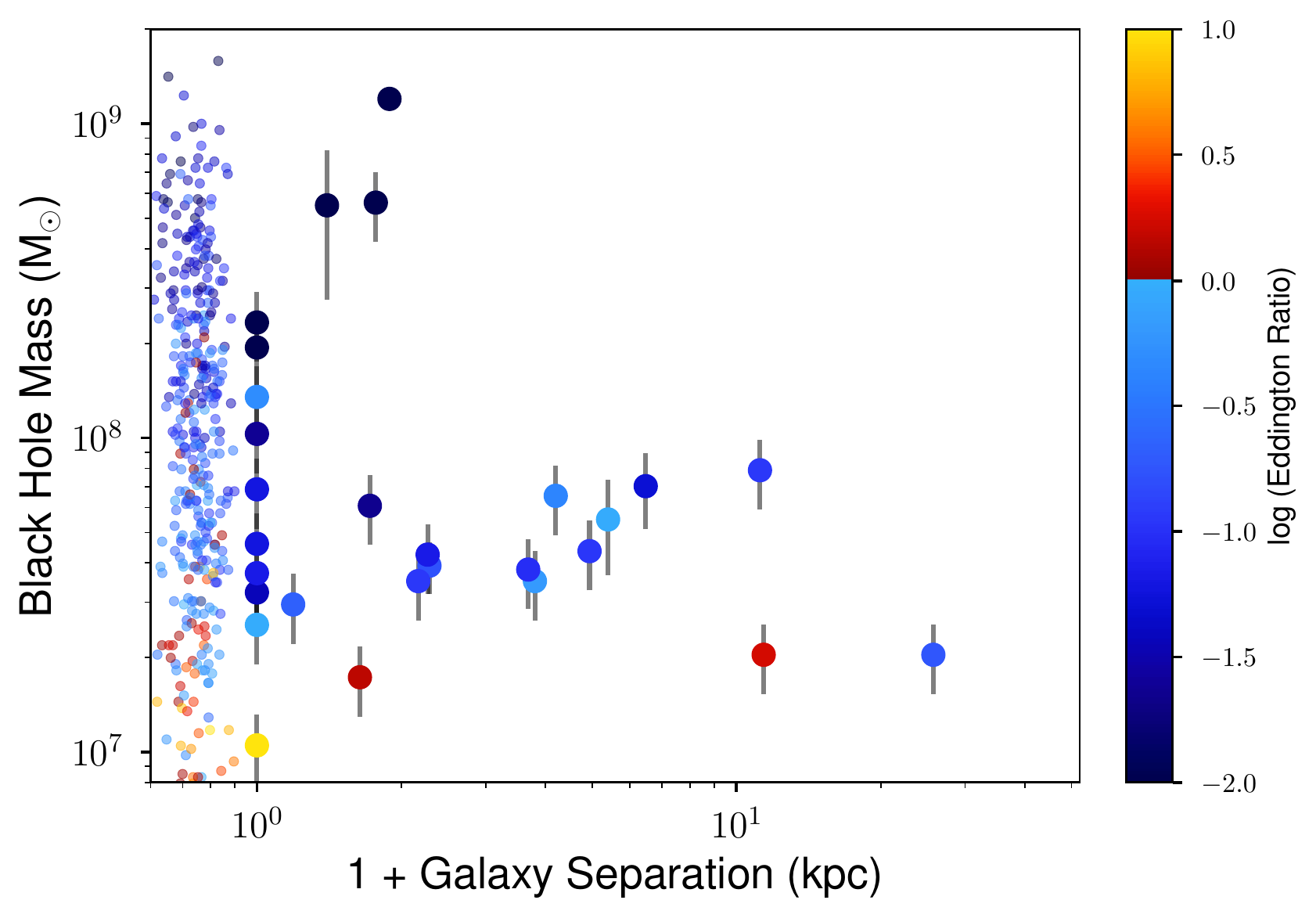}
\caption{Figures \ref{fig:AGNbh} lower and \ref{fig:merge_agn1}, made using fits that replace the CYGNUS AGN models with the \citealt{stal16} clumpy AGN models.}
\label{fig:altexski}
\end{center}
\end{figure*}

\section{Alternative Explanations}\label{app:alt}
The results in \S\ref{bhmass} depend on both a model-based anisotropy correction, and the choice of black hole mass measurements. We here explore the robustness of the results against these choices. 

\subsection{Choice of radiative transfer model}\label{appradtrans}
 We adopt the CYGNUS-based results from EFS21, as EFS21 show they provide the best fits to the photometric and spectroscopic observations of our sample, on average, especially at mid-infrared wavelengths. EFS21 do however present results using three other AGN models, from \citet{sieben15}, \citet{fritz06}, and \citet{stal16}. As described in EFS21, these four AGN models make significantly different physical assumptions, and can give different component SED shapes for the same source.  Comparing the results obtained in this paper from using all four AGN models thus gives a reasonably robust test of our results against the choice of AGN model. 

The \citealt{fritz06} AGN model gives conceptually similar results to the CYGNUS AGN  model in both the $\rm{M}_{BH} - \lambda_{e}$ and $\rm{M}_{BH}  - n_{s}$ planes (\ref{fig:altexfri}). With the \citealt{sieben15} AGN model (Figure \ref{fig:altexsie}) the fitted slope to the super-Eddington systems in the $\rm{M}_{BH} - \lambda_{e}$ is flatter than with the CYGNUS models, and virtually identical to the slope for the sub-Eddington systems. The results in the $\rm{M}_{BH}  - n_{s}$ are however unchanged. When the Stalevski AGN model (Figure \ref{fig:altexski}) is used, few objects harbour super-Eddington AGN, and the results obtained with the CYGNUS models are effectively absent. However, as shown in EFS21, the Stalevski AGN model does not provide as good fits to the SEDs of our sample as do the other AGN models.

\begin{figure*}
\begin{center}
\includegraphics[width=7cm]{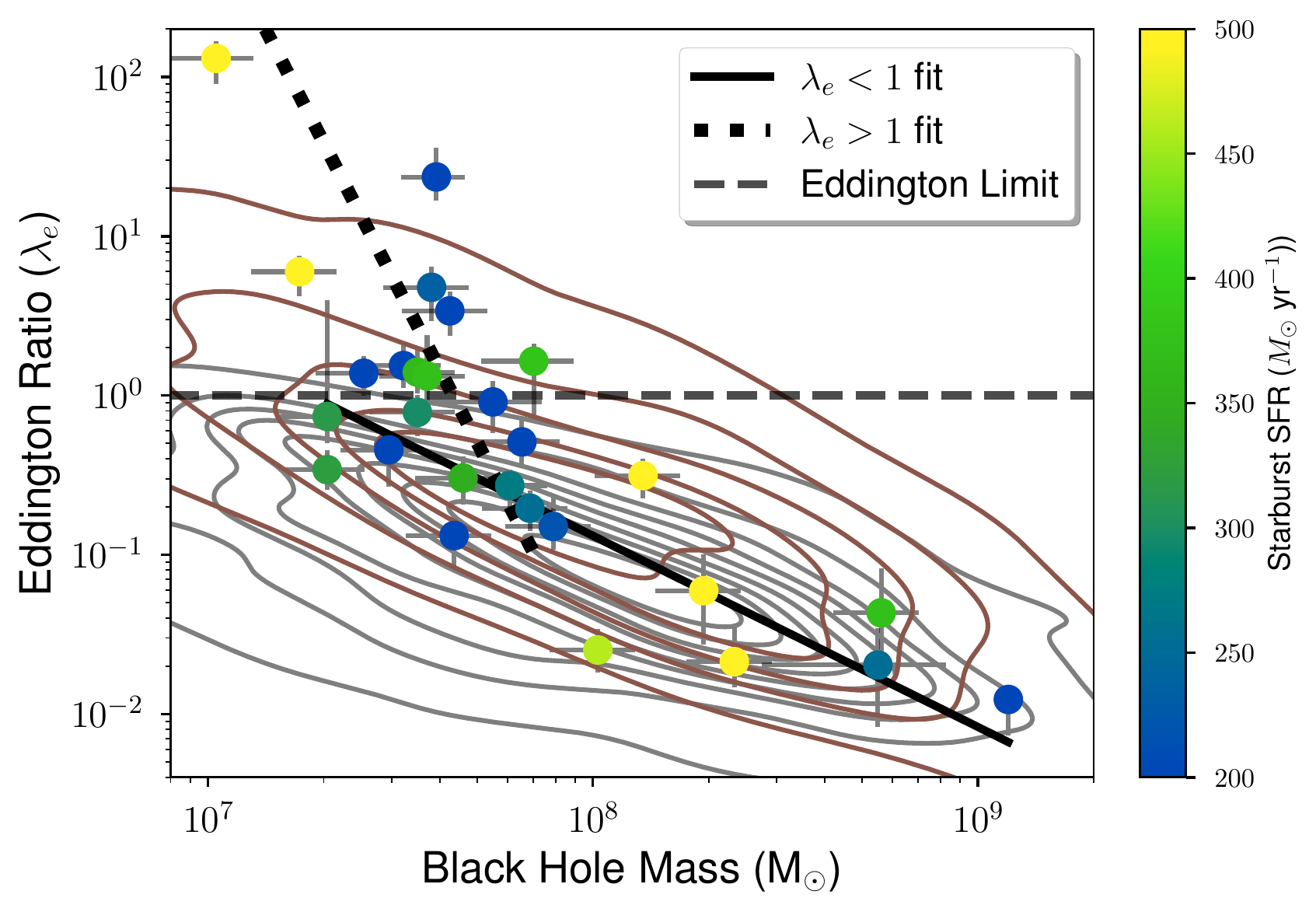}
\includegraphics[width=7cm]{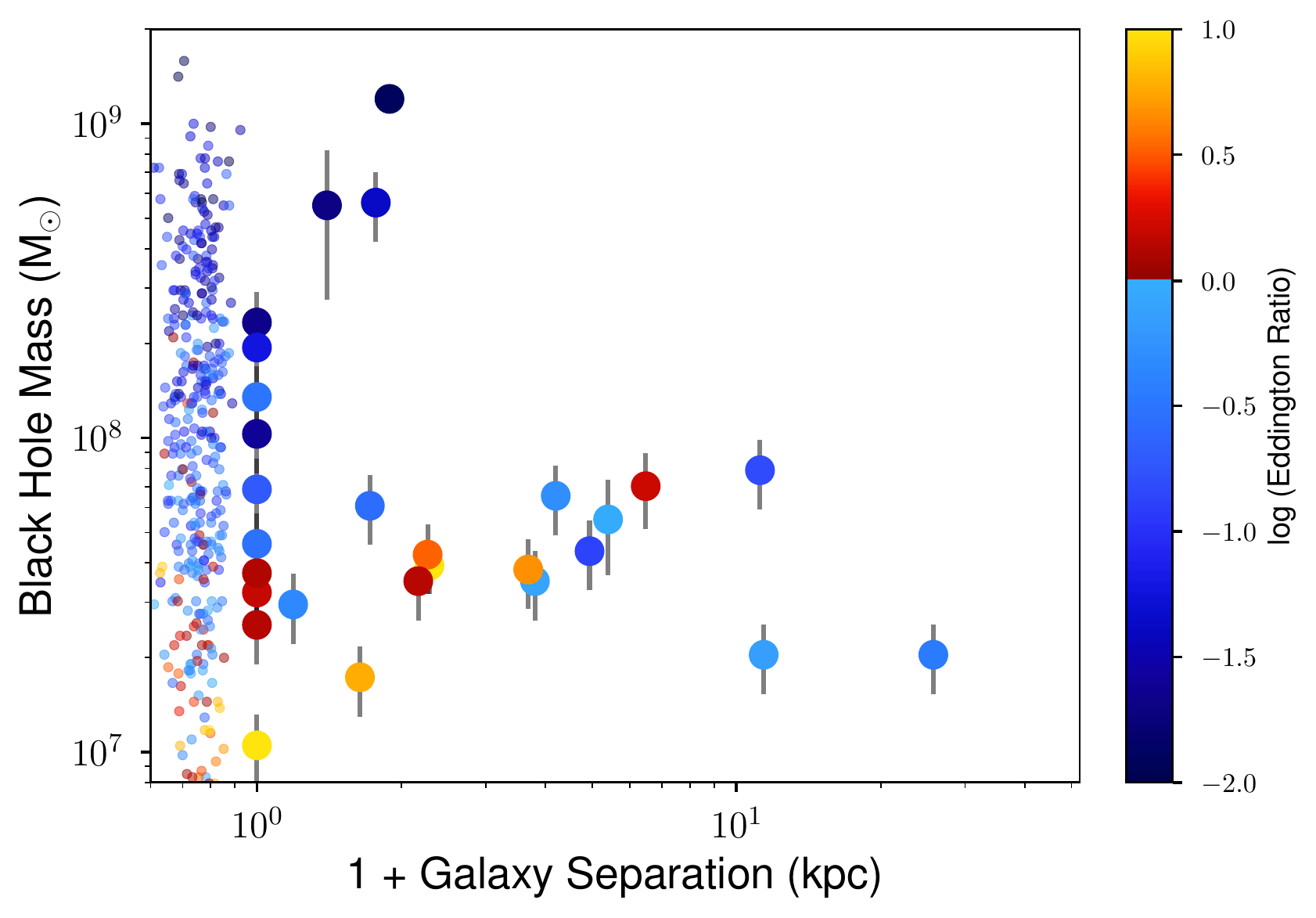}
\caption{Figures \ref{fig:AGNbh} lower and \ref{fig:merge_agn1}, made using the fits with the highest maximum likelihoods for each object. This leads to a mix of AGN models from all four used in EFS21. The results are almost identical to those from using just the CYGNUS model fits.}
\label{fig:altexbest}
\end{center}
\end{figure*}

A final option is to adopt the fit results with the highest maximum likelihoods for each object, irrespective of the AGN model used. This gives the best possible fit for each object, but also a mix of mostly CYGNUS and Siebenmorgen et al results, with a few using the Fritz models and one using the Stalevski models. The results  (Figure \ref{fig:altexbest}) are virtually identical to those using the CYGNUS models alone. 

We conclude from this analysis that our results are reasonably robust to the adopted AGN model.

\begin{table}
\begin{center}
\begin{tabular}{ccc}
\hline 
ID & $\rm{M_{BH}}$       & $\rm{M_{BH}}$    \\ 
   & Photometric         & Virial           \\
   & \multicolumn{2}{c}{$10^{7}\rm{M_{\odot}}$} \\
\hline 
1  & 26.6  & ---    \\     
2  & 31.5  & ---    \\          
3  & 7.61  & ---    \\         
5  & 28.9  & ---    \\         
6  & 20.8  & ---    \\         
9  & 39.0  & 18.0   \\                   
11 & 9.00  & ---    \\                 
13 & 19.7  & 1.02   \\    
15 & 7.53  & ---    \\           
16 & 27.4  & ---    \\             
18 & 71.7  & ---    \\    
19 & 40.7  & ---    \\                          
22 & 11.4  & 0.24   \\    
23 & 37.0  & ---    \\          
28 & ---   & 390.0  \\    
29 & 10.7  & ---    \\          
31 & 27.1  & ---    \\         
37 & 37.8  & 9.81   \\   
38 & 19.9  & ---    \\          
40 & 15.1  & ---    \\        
42 & 125.8 & 9.05   \\   
\hline	
\end{tabular}
\caption{The available photometric and single-epoch virial black hole masses for our sample. These data are taken from \citealt{greene07,kawa07,vei09,medling15,harv16}. Only objects with at least one of a virial or photometric black hole mass are listed.}
\label{tablephotvirbh}
\end{center}
\end{table}

\subsection{Choice of black hole mass estimates}\label{appbhmass}
Three types of black hole mass estimate are available for subsets of our sample; dynamical (28 objects), photometric (20 objects), and single-epoch virial (6 objects). The reliability of all three measures in low-redshift infrared-luminous mergers has been previously discussed, but without a preference emerging \citep{dasyra06,vei09}. This is in part because virtually no reverberation-mapping black hole mass estimates exist for infrared-luminous mergers to compare to, and very few objects in which the three estimates can be intercompared. Nevertheless, we here examine the consequences of using photometric or virial, rather than dynamical black hole masses. For convenience, we summarize these masses in Table \ref{tablephotvirbh}.

\begin{figure*}
\begin{center}
\includegraphics[width=7cm]{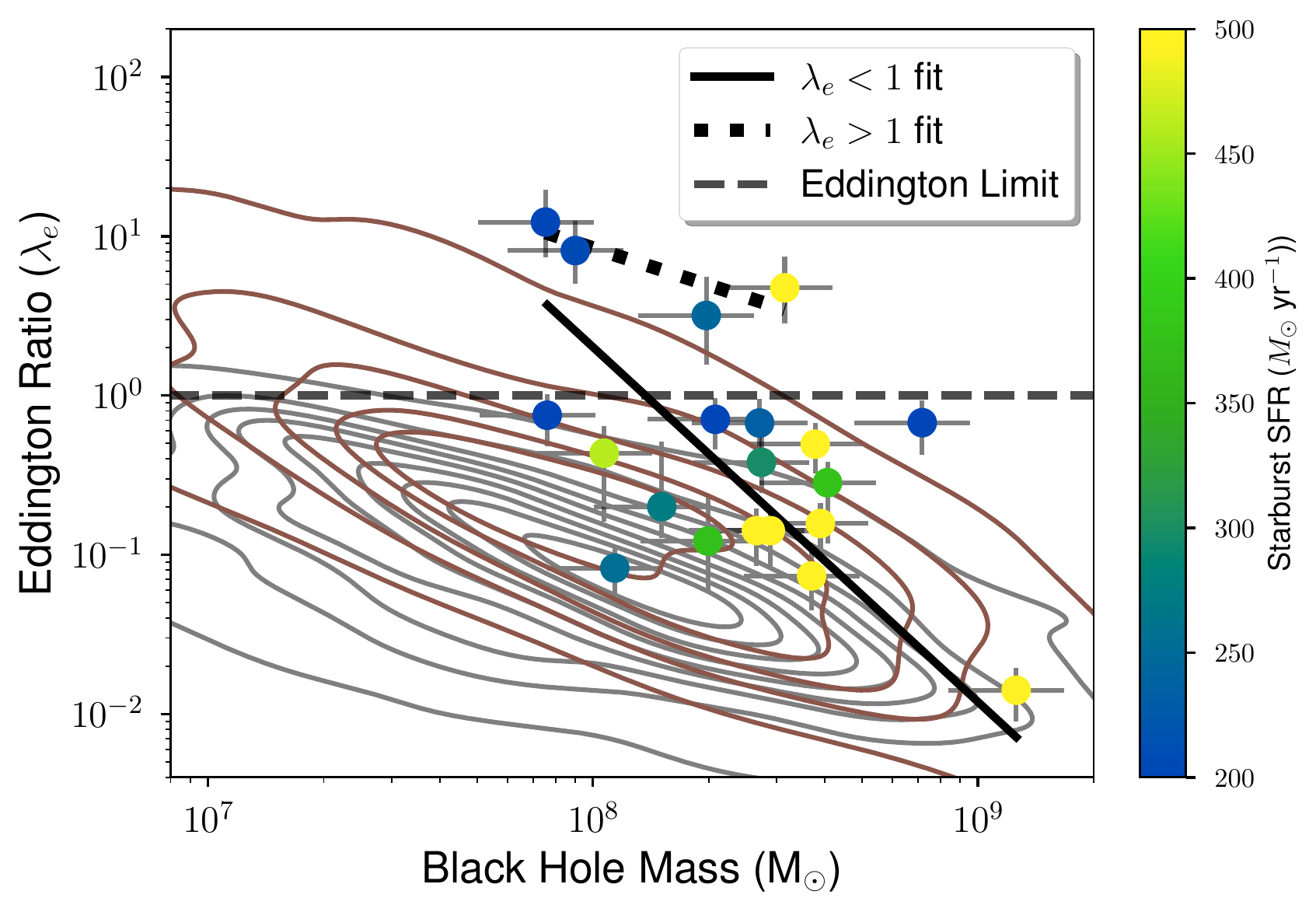}
\includegraphics[width=7cm]{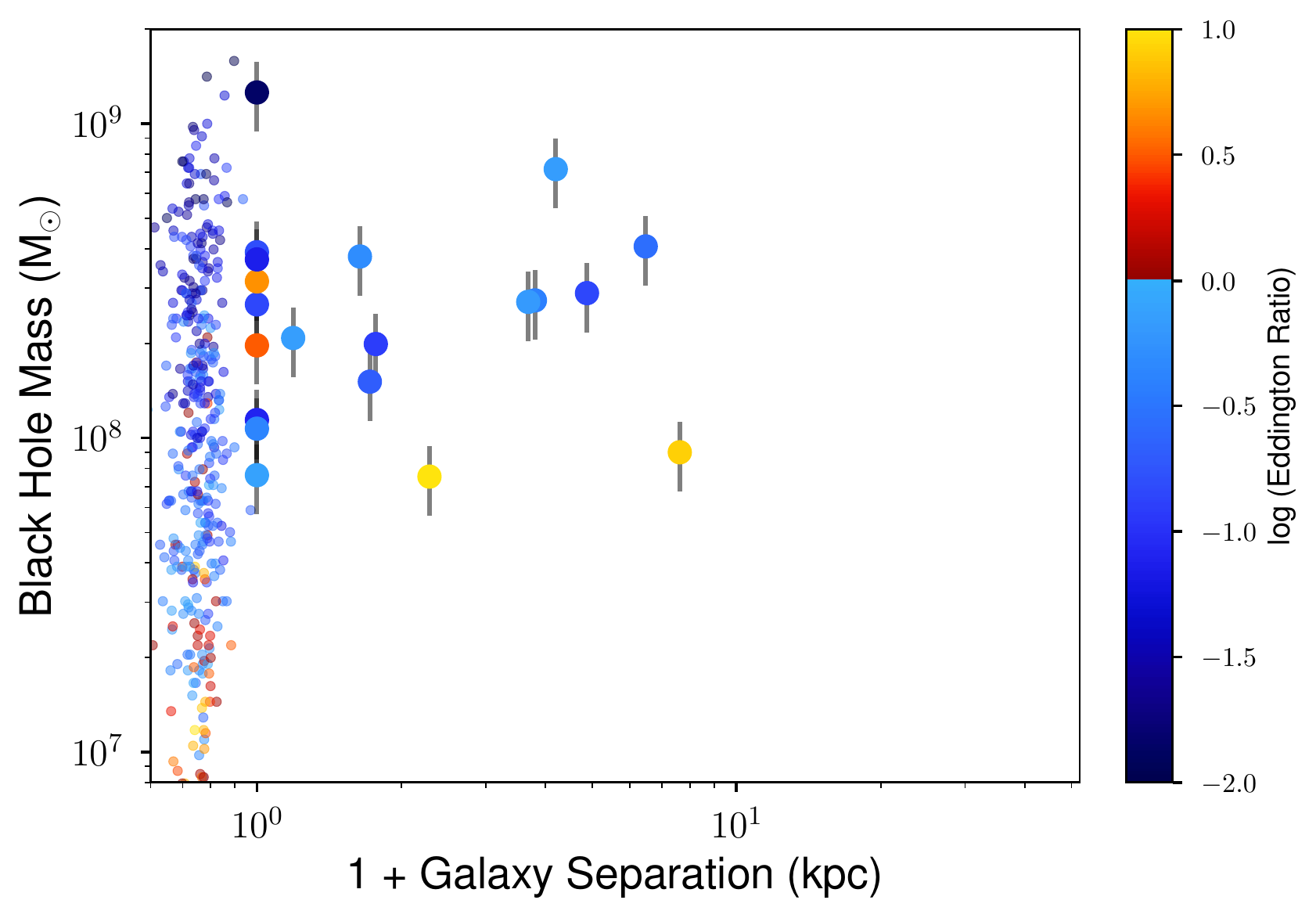}
\caption{Figures \ref{fig:AGNbh} lower and \ref{fig:merge_agn1}, made by replacing dynamical black hole masses with photometric black hole masses \citep{vei09}. Using these masses, few systems harbour super-Eddington AGN and there is no evidence for an association with late-stage mergers.}
\label{fig:altexusephoto}
\end{center}
\end{figure*}

The photometric masses are usually several times larger than the dynamical masses. Using the photometric masses  in our analysis instead of the dynamical masses effectively eliminates the results obtained using the dynamical masses; there are far fewer objects with evidence for super-Eddington accretion, and they show no evidence for association with late-stage mergers (Figure \ref{fig:altexusephoto}). 

\begin{figure*}
\begin{center}
\includegraphics[width=7cm]{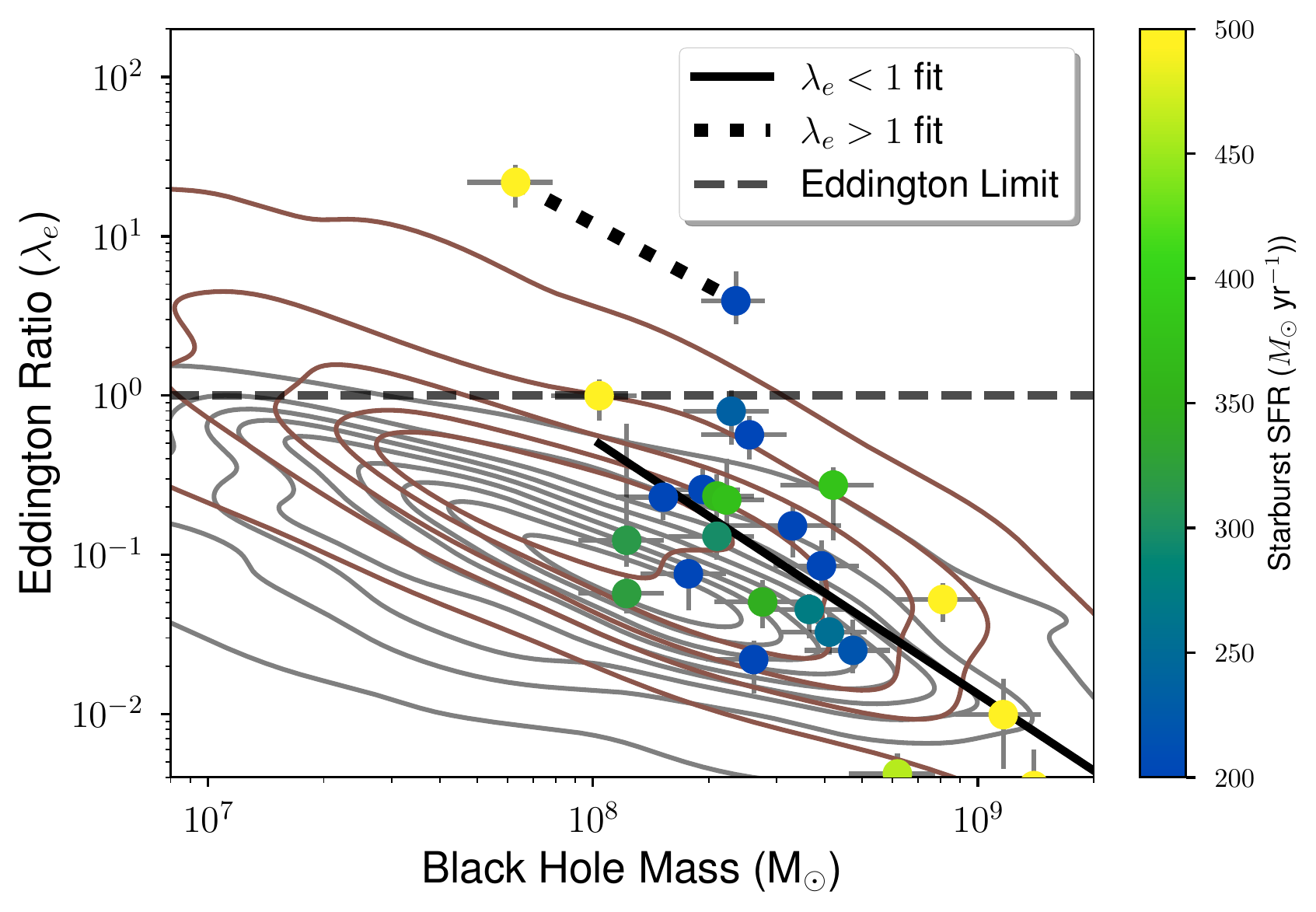}
\includegraphics[width=7cm]{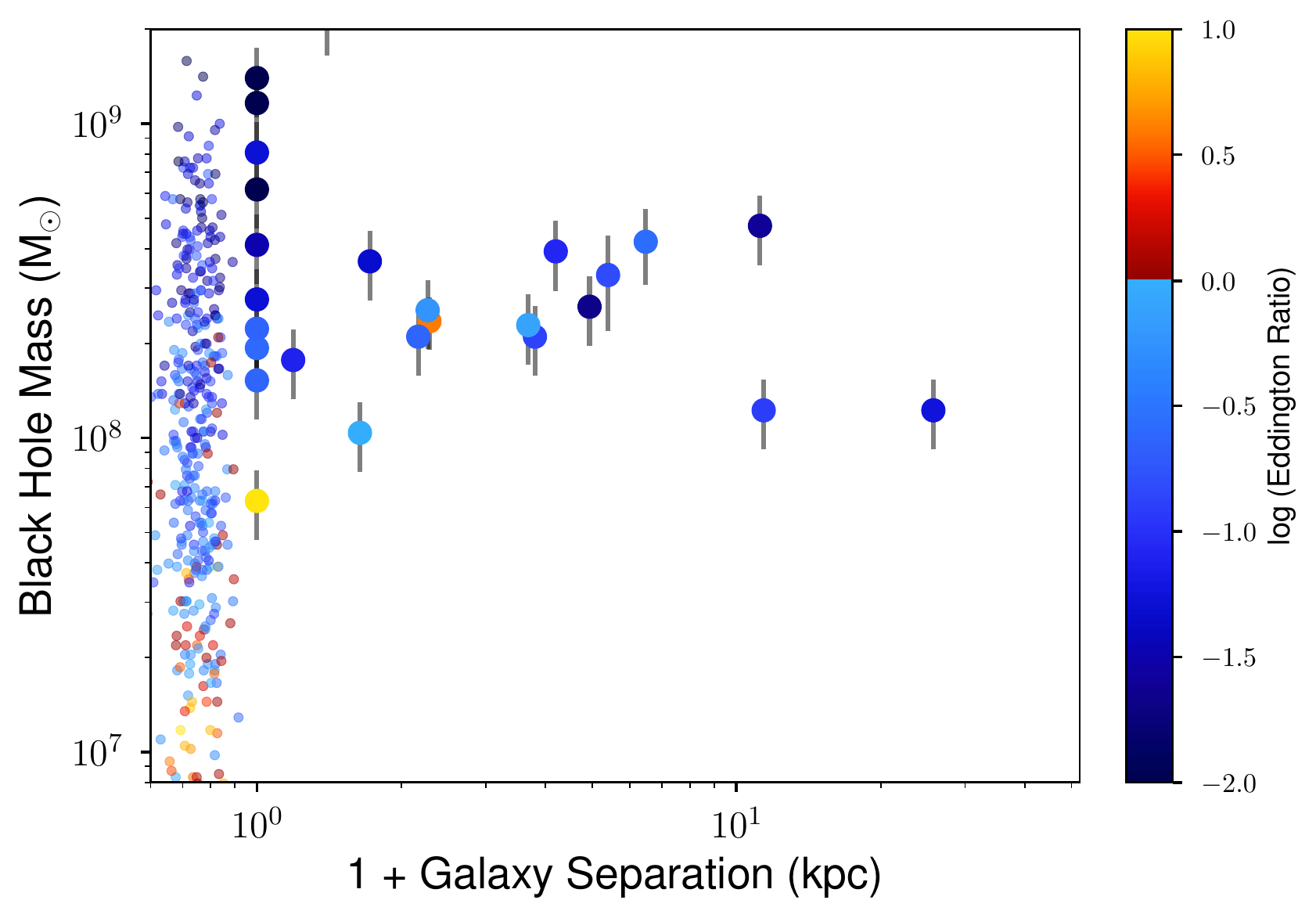}
\caption{Figures \ref{fig:AGNbh} lower and \ref{fig:merge_agn1}, made by incrementing the dynamical black hole masses by a factor of six. The same results are (just) visible.}
\label{fig:altexsys}
\end{center}
\end{figure*}

\begin{figure*}
\begin{center}
\includegraphics[width=7cm]{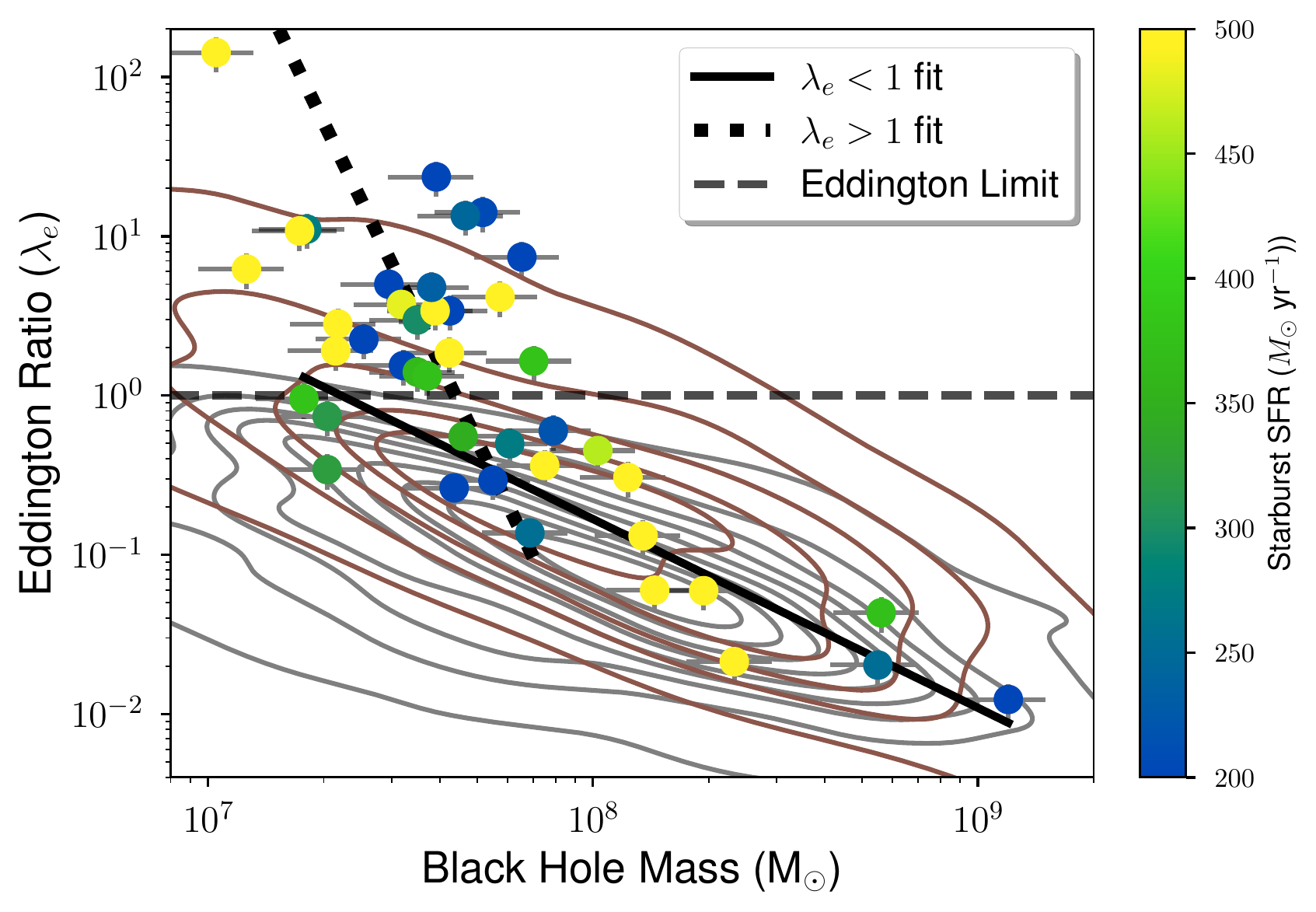}
\includegraphics[width=7cm]{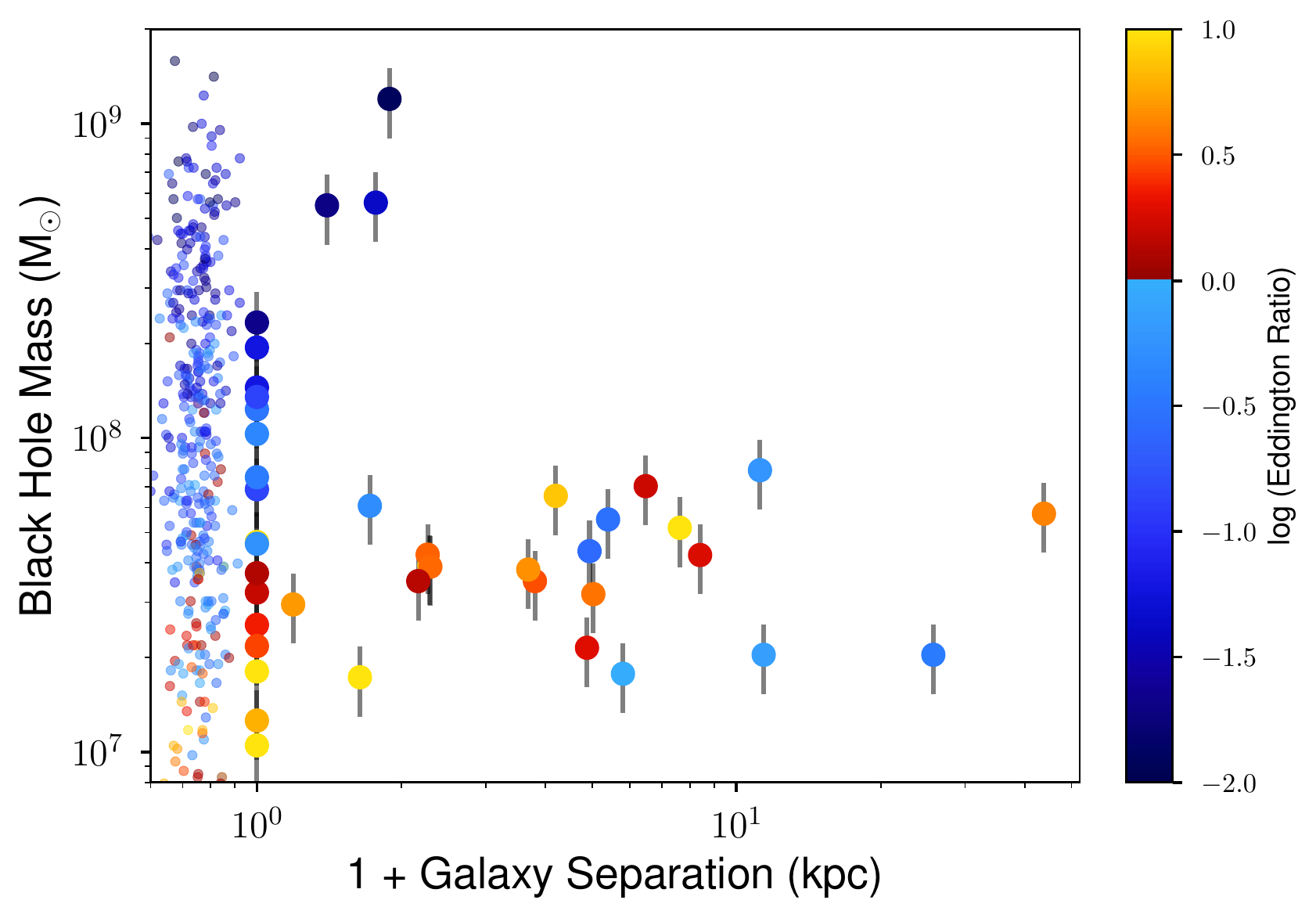}
\caption{Figures \ref{fig:AGNbh} lower and \ref{fig:merge_agn1}, made by  generating artificial black hole masses for the 14 objects without dynamical masses (by drawing from a uniform distribution over $\log{\rm{M}_{BH}} = 7.1 - 8.4$) and adding them to both panels. The results are unchanged in both cases.}
\label{fig:altexfake}
\end{center}
\end{figure*}

We propose however that the photometric black hole masses are less likely to be accurate in our sample than the dynamical masses. Adopting photometric black hole masses clearly offsets the sample from both the type 1 and type 2 AGN comparison samples.  If we assume that our sample are a subset of the  local type 2 AGN population, and are the antecedents of type 1 QSOs,  then the relative positions of the sample to these populations in the left panel of Figure \ref{fig:altexusephoto} would be challenging to explain. The dynamical masses on the other hand make the sample consistent with both hypotheses. We conclude that the photometric black hole masses are in general overestimated, and that the dynamical black hole masses are more likely to be closer to the true values. 

There are few virial mass estimates for our sample and nearly all are consistent with the dynamical mass estimates. Including the virial masses for objects that do not have a dynamical mass does not change the results, and may complicate their interpretation. So, we do not use them so as to maintain a single mass estimation method. 

\subsection{Systematics in black hole mass estimates}\label{appsystem}
Assuming the dynamical masses are the best available, it is possible that they are systematically biased in some way. We here consider the consequences for our results of such a bias.

First, it has been proposed that the dynamical masses of black holes in local elliptical galaxies are biased high with respect to the true average, as the dynamical approach relies on resolving the black hole sphere of influence \citep{shan16}. This effect, if real, is unlikely to affect our results as our sample includes most local infrared-luminous mergers, rather than a small subset. Even if it did, it would act to increase the number of super-Eddington systems, rather than decrease them. 

A second possibility is that the black hole masses are biased to low values. We are not aware of any evidence for such a bias, but given the paucity of comparisons between dynamical and reverberation masses, we consider it here.  Systematically increasing the dynamical black hole masses gives essentially the same results, until the increase reaches a facor of about seven, at which point the results are significantly affected (Figure \ref{fig:altexsys}).   We conclude that, even if a systematic bias to low values is found in dynamical black hole masses, it will not affect our results unless it exceeds a factor $\sim6$.
 
\subsection{Black hole mergers}\label{appmergers}
The $M_{BH}$ increase in late-stage mergers could in part be due to an SMBH-SMBH merger. This is an effect we account for, but we discuss it further here. 

The total mass increase we infer however is around an order of magnitude, while it is hard for a black hole merger to contribute a mass increase substantially in excess of a factor of two (doing so requires assuming a greater preference for the AGN to be the lighter black hole). It is also worth noting that a black hole merger timescale of much less than $10^{8}$ years is controversial, with some claims it is possible \citep{esca05,colpi14,khan16}, others not \citep{tamb17}.

\subsection{Sample bias}\label{abias}
Fourteen of our sample do not have dynamical black hole mass measurements. It is in principle possible that these fourteen sources could dilute or eliminate our results, if such mass measurements were available for them. 

Two of these fourteen objects (\#9 \& 28) have optical- or OH-based virial black hole mass estimates; $1.8\times10^{8}$M$_{\odot}$ \citep{kawa07} and $3.9\times10^{9}$M$_{\odot}$ \citep{harv16}. Adopting these masses maintains consistency with our original result. 

Six of the remaining twelve systems (\# 1, 10, 13, 17, 23, 34) are single-nucleus. Adopting random black hole masses between the minimum and maximum seen in the sample does not significantly change the result. 

The other six systems (\# 4, 5, 8, 11, 12, 26) have double nuclei. Five of them are relatively close separation systems, and whether or not none, some, or all are super-Eddington would not substantively change our result. One system is a wide separation system with a luminous AGN. Assuming the separation is accurate, then if this sytem were super-Eddington it would dilute our result slightly, but we have no way to test this. 

To illustrate, we take all fourteen objects without a dynamical black hole mass, and assign each of them an artificial black hole mass drawn from a uniform distribution spanning $\log{\rm{M}_{BH}} = 7.1 - 8.4$. In 100 realizations of such an experiment, our results are essentially unchanged. An example is shown in Figure \ref{fig:altexfake}.

\subsection{Where are the super-Eddington type 1 Quasars?}\label{appsequso}
If a significant fraction of our sample exhibit super-Eddington accretion, then it might be expected that some type 1 AGN should also show super-Eddington accretion. 

The fraction of type 1 super-Eddington AGN is uncertain, but likely very small; less than $\sim3\%$ of the type 1 quasars in SDSS DR7 have $\lambda_{e}>1$ (\citealt{shen11}, see also \citealt{ves09}) and most other candidate super-Eddington systems are narrow-line AGN (see main text for references). Such a small fraction is however not in tension with our results. Several independent lines of evidence imply high covering fractions at high accretion rates, including e.g. Figure \ref{fig:AGNbh} of this work, high hydrogen columns in highly accreting AGN \citep[e.g.]{ricci21}, and several theoretical lines of evidence (summarized in \S\ref{hidpop}). It is thus plausible that the super-Eddington AGN have far fewer lines of sight to the broad-line region, leading to few observed type 1 super-Eddington AGN.

\subsection{Inclination angle bias}\label{appinclbias}
A bias in the inclination angle of the AGN torus relative to the observer is unlikely to affect our results. There is no correlation between inclination angle and corrected AGN luminosity. Furthermore, the CYGNUS inclination angles are broadly consistent with those measured using other methods (EFS21). The sample also contains both obscured and unobscured AGN spectral types, independently implying a range in inclination angle. 

\subsection{Transient Phenomena}\label{apptransi}
A population of transients, e.g. AGN flaring, nuclear supernovae, or tidal disruption events, are unlikely to be causing objects to appear super-Eddington. The data used in the modelling were taken over timescales of over ten years. They also sample emission from spatial scales of tens of parsecs. Since transients are bright for at most a few years, it seems unlikely that they could systematically affect our results. Moreover, the modelling approach used in EFS21 has previously successfully identified at least some of these transient phenomena \citep{mat18}.

\subsection{Host obscuration degeneracy}\label{apphostobs}
As discussed in e.g. EFS21, there is a degeneracy in multi-component model fits; if the optical depth of the host galaxy model is allowed to reach arbitrarily high values then the luminosity of the AGN component can be raised to extreme values to compensate. The host model in the fits is however limited to reasonable bounds, based on comparisons to low-redshift disk galaxies (EFS21) so this effect is unlikely to bias our results. 

\subsection{Deviation from Thomson Scattering}\label{appthomdev}
The typical opacity may be lower than the Thomson scattering value, but this appears unlikely as an alternative explanation for our results since most alternative scenarios have higher opacities.

\bsp
\label{lastpage}
\end{document}